\documentclass[pra,aps,amssymb,twocolumn,noshowpacs]{revtex4-1}
\usepackage{color}
\usepackage{graphicx}
\usepackage{hyperref}
\usepackage{amsmath}
\usepackage{latexsym}
\usepackage{amssymb}
\usepackage{mathrsfs}
\usepackage{mathtools}
\usepackage{layout}
\usepackage{verbatim}
\usepackage{bm}
\usepackage{amsfonts,epsfig}

\usepackage{tikz}

\begin{document}

\title{Fast, Accurate, and Realizable Two-Qubit Entangling Gates by Quantum Interference in Detuned Rabi Cycles of Rydberg Atoms}

\date{\today}
\author{Xiao-Feng Shi}
\affiliation{School of Physics and Optoelectronic Engineering, Xidian University, Xi'an 710071, China}

\begin{abstract} 
High-fidelity entangling quantum gates based on Rydberg interactions are required for scalable quantum computing with neutral atoms. Their realization, however, meets a major stumbling block -- the motion-induced dephasing of the transition between the ground and Rydberg states. By using quantum interference between different detuned Rabi oscillations, we propose a practical scheme to realize a class of accurate entangling Rydberg quantum gates subject to a minimal dephasing error. We show two types of such gates, $U_{1}$ and $U_{2}$, in the form of $\text{diag}\{1, e^{i\alpha}, e^{i\gamma}, e^{i\beta}\}$, where $\alpha, \gamma$, and $\beta$ are determined by the parameters of lasers and the Rydberg blockade $V$. $U_{1}$ is realized by sending to the two qubits a {\it single} off-resonant laser pulse, while $U_{2}$ is realized by individually applying one pulse of detuned laser to each qubit. One can construct the controlled-z~(C$_{\text{Z}}$) gate from several $U_{1}$ when assisted by single-qubit rotations and phase gates, or from two $U_{2}$ and two phase gates. Our method has several advantages. First, the gates are accurate because the fidelity of $U_k$ is limited only by a rotation error below $10^{-5}$ and the Rydberg-state decay. Decay error on the order of $10^{-5}$ can be easily obtained because all transitions are detuned, resulting in small population in Rydberg state. Second, the motion-induced dephasing is minimized because there is no gap time in which a population is left in the Rydberg shelving states of either qubit. Third, the gate is resilient to the variation of $V$. This is because among the three phases $\alpha,\gamma$, and $\beta$, only the last has a (partial) dependence on $V$. Fourth, the Rabi frequency and $V$ in our scheme are of similar magnitude, which permits a fast implementation of the gate when both of them are of the feasible magnitude of  several megahertz. The rapidity, accuracy, and feasibility of this interference method can lay the foundation for entangling gates in universal quantum computing with neutral atoms.

\end{abstract}
\maketitle

\section{introduction}
Ultracold atoms are promising for scalable quantum computing~\cite{Saffman2010,Saffman2016,Weiss2017} due to the ability to isolate and detect single atoms, to trap atoms in arrays of optical dipole traps, to store quantum information in the hyperfine sublevels of the ground state, and to perform high-fidelity single-qubit gates with individual addressing~\cite{PhysRevLett.103.153601,RSI2014,Nogrette2014,Xia2015,Zeiher2015,Ebert2015,Wang2016,Barredo2016}. In order to develop a reliable quantum computer of practical utility, a universal set of accurate, fast, and realizable quantum gates are required. Any two-qubit entangling quantum gate and a small number of single-qubit gates can form a universal set of quantum gates~\cite{PhysRevLett.89.247902}. This leads to the general conclusion that a high-fidelity entangling gate is necessary for the development of scalable quantum computing. However, despite the fact that two-qubit gates with ultracold neutral atoms based on Rydberg interactions~\cite{Gallagh2005} were proposed about two decades ago~\cite{PhysRevLett.85.2208}, and in theory high fidelity is achievable~\cite{Goerz2014,Theis2016,Shi2017,Petrosyan2017,Shi2018prapp2}, there is no substantial progress in recent experiments~\cite{Wilk2010,Isenhower2010,Zhang2010,Maller2015,Jau2015,Zeng2017,Picken2018}.  

A major factor that leads to the low fidelity of Rydberg quantum gates is the motion-induced Doppler dephasing of the atomic transition between the qubit states and Rydberg states~\cite{Wilk2010,Saffman2011,Saffman2016}. To date, all Rydberg quantum gates~\cite{Isenhower2010,Zhang2010,Maller2015,Zeng2017} were experimentally studied using the three-pulse method originally proposed in~\cite{PhysRevLett.85.2208}, where the control qubit is left in a Rydberg shelving state when the target qubit is optically pumped. A problem with this method is that when the qubits drift, the phases of the laser fields used for excitation and deexcitation of the control qubit can have sizable difference. Such Doppler dephasing results in both population and phase error in the gate. This dephasing can be suppressed if no gap time is left between the excitation and deexcitation of Rydberg states. In other words, the motion-induced dephasing can be minimized if each qubit is driven only by one pulse of laser field~\cite{DeLeseleuc2018}. This is why two-atom entanglement between ground and Rydberg states was only recently achieved with a high fidelity by applying a single laser pulse~\cite{Levine2018}. However, it is the entanglement between ground-state atoms that is useful for scalable quantum computing, but a practical dephasing-resilient scheme for this goal is still lacking.

\begin{table*}[ht]
  \centering
  \begin{tabular}{lcccc|cccc}
    \hline
    \hline   \text{Case}&     $ \frac{\Omega}{2\pi}$~(MHz)&  $\frac{ \Delta}{2\pi}$ (MHz)  &$\frac{  V}{2\pi} $ (MHz) &  $\frac{\beta-2\alpha}{\pi} $& $ (N,M_1,M_2,M_3) $  &$t_{\text{g}}$ (ns) &  $E_{\text{ro}}$& $E_{\text{de}} $\\ \hline
1&	 10  & 19.252&  -35.1818 &  0.32457&(4,~2,~1,~-3)  &184 &$2.31\times10^{-10} $&  45.5ns/$\tau$ 	 \\ 
2&	 10  & -23.9977& 52.1713& 0.6217118& (10,~8,~-3,~-5)   & 385&$6.80\times10^{-9} $&  59.9ns/$\tau$ 	 \\ 
3&	 10  & -13.6468& 23.09272&  1.450098 &(5,~4,~-1,~-3)  &296 &$3.59\times10^{-8}$ &  86.9ns/$\tau$ 	 
\\ \hline \hline
  \end{tabular}
  \caption{  \label{table1} Three sets of parameters $N, \Omega, \Delta$, and $V$ for realizing $U_1$ that satisfy Eq.~(\ref{interference01})~[or Eq.~(\ref{gatetime01})] exactly, and simultaneously satisfy Eq.~(\ref{interference})~ [or Eq.~(\ref{gatetime02})] approximately with errors small than $10^{-7}$. The angle $\beta-2\alpha $ shown here is deducted by an multiple of $2\pi$ so that it falls into the interval $[0,2\pi)$. Many digits for $ \Delta$ and $V$ are displayed because they are from numerical optimization. Here an arbitrary phase can be added to $\Omega$ with the same results achieved here. Note that for each case, a similar gate can be realized with the angle $-(\beta-2\alpha)$ when the signs of both $\Delta$ and $V$ are reversed.}
  \end{table*}


\begin{figure*}[ht]
  \centering
\includegraphics[width=5in]
{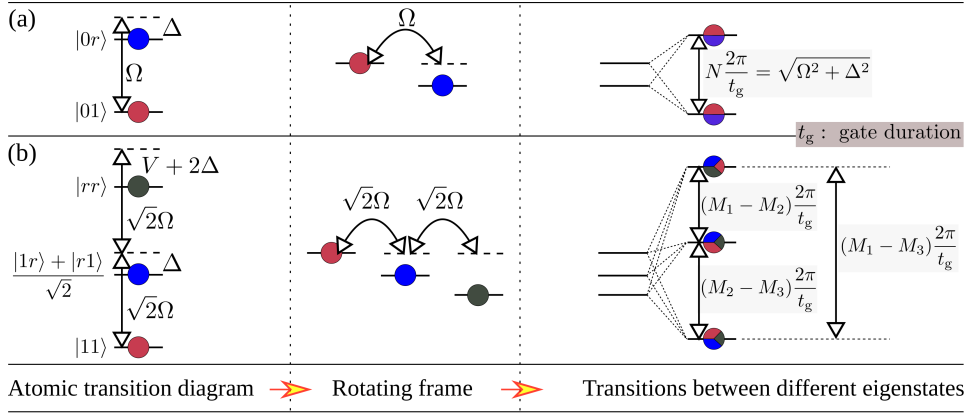}
 \caption{(a) and (b) show how rational generalized Rabi frequencies in $U_1$ appear for the two input states $|01\rangle$ and $|11\rangle$, respectively. Here ``rational'' refers to that they are characterized by integers $(N,M_1,M_2,M_3)$. A picture similar to (a) also applies to the input state $|10\rangle$. After transformed to the rotating frame, the Hamiltonian for the atomic transition is diagonalized. The energy difference~(divided by the reduced Planck constant $\hbar$) between the two eigenstates on the rightmost of (a) is $\sqrt{\Omega^2+\Delta^2}$, which is equal to $N$ times of $2\pi/t_{\text{g}}$, where $t_{\text{g}}$ is the duration of the gate sequence. For the input state $|11\rangle$, the detuned optical pumping and the Rydberg blockade lead to three Stark-shifted eigenstates. Between these states there are three transition frequencies, as shown in the three shaded boxes on the rightmost of (b). The entangling gate $U_1$ can be achieved when the three transition frequencies are given by $(M_i-M_j)2\pi/t_{\text{g}}$ with three different integers $M_1,~M_2$, and $M_3$, where $i\neq j$ and $i,j\in1-3$,  \label{figure01new} }
\end{figure*}

\begin{figure}
\includegraphics[width=3.0in]
{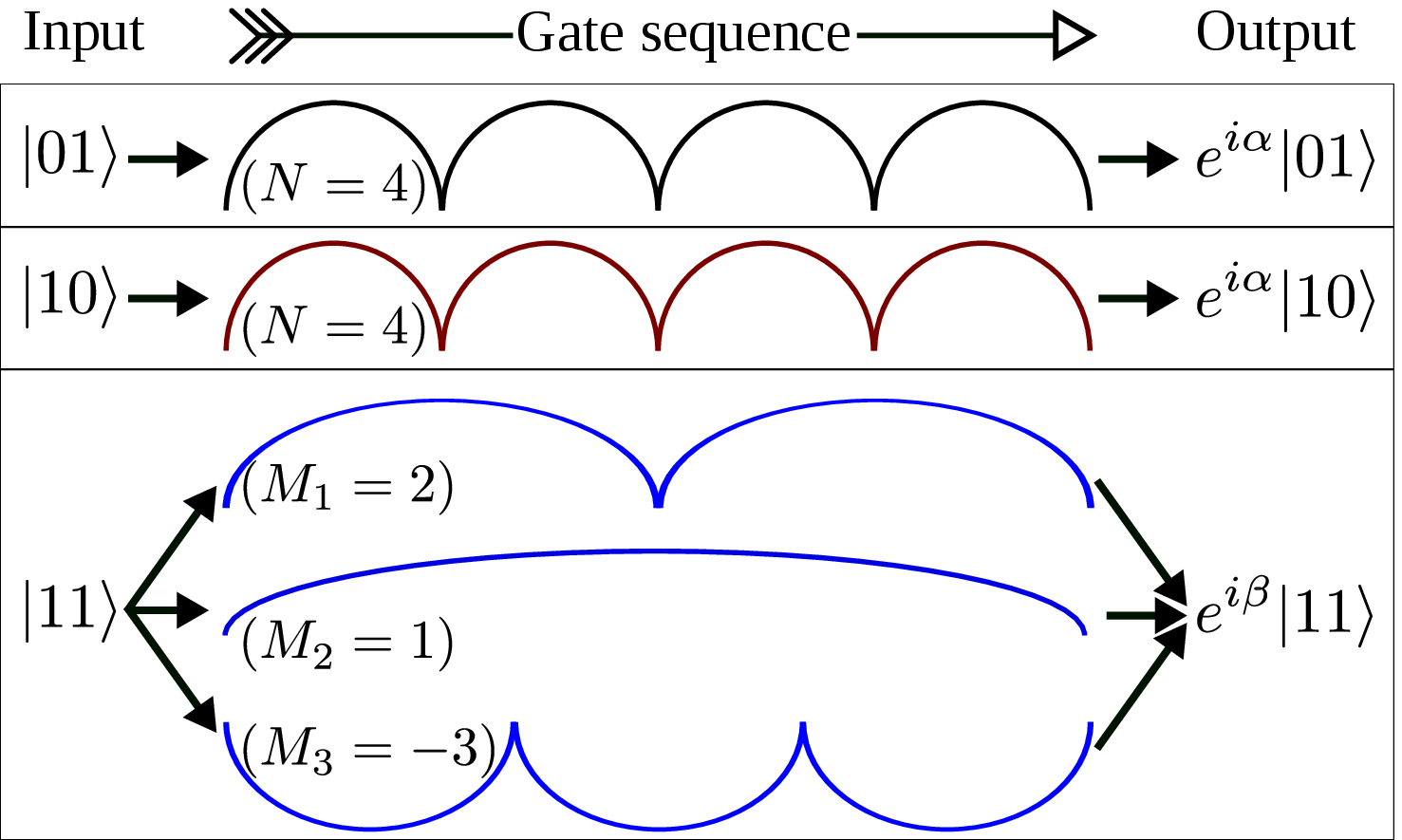}
 \caption{Illustration of the quantum interference method to realize the $U_1$ gate with parameters of case 1 in Table~\ref{table1}. The input states $|01\rangle$ and $|10\rangle$ undergo four generalized Rabi cycles~($N=4$) which imprints a phase $\alpha$ upon them, see Eq.~(\ref{interference01})~[or Eqs.~(\ref{gatetime01}), (\ref{trans01}),  and~(\ref{alpha})]. For the input state $|11\rangle$, both laser detuning and vdWI block its resonance which results in three types of transition frequencies between them. Then, it undergoes three types of detuned Rabi oscillations characterized with $(M_1,M_2,M_3)=(2,1,-3)$, see Eq.~(\ref{interference}) and Fig.~\ref{figure01new}. When the laser excitation completes, the three off-resonant Rabi oscillations in $|11\rangle$ also finish; this is equivalent to that the three quantum evolution pathways interfere constructively so that the population is restored to $|11\rangle$. The overall effect of these three detuned Rabi cycles is a pure phase change of $\beta$ for the input state $|11\rangle$.  \label{figure01} }
\end{figure}

In this work, we propose two types of high-fidelity two-qubit entangling gates $U_{1(2)}$ which are resilient to the motion-induced Doppler dephasing. Our method is based on quantum interference between different detuned Rabi oscillations~\cite{Shi2017,Shi2018prapp2}, and is subject to a minimal Doppler dephasing because it is implemented either with a single square laser pulse on two qubits~(for $U_1$), or with one square laser pulse on each qubit~(for $U_2$). As a quick guide to the essence of this method, we show the patterns of the atomic transition and quantum interference of $U_1$ in Fig.~\ref{figure01new} and Fig.~\ref{figure01}, respectively. Figure~\ref{figure01new} shows that there are two Stark-shifted states for the input state $|01\rangle$, between which there is a transition frequency $2\pi N/t_{\text{g}}$, where $t_{\text{g}}$ is the gate duration. For the input state $|11\rangle$, there are three Stark-shifted eigenstates separated by three transition frequencies $(M_i-M_j)2\pi/t_{\text{g}}$, where $i\neq j$ and $i,j\in1-3$. For any set of integers $\{N,M_1,M_2,M_3\}$ that satisfy the physical picture of Fig.~\ref{figure01new}, the Rabi frequency for $|01\rangle$~(or $|10\rangle$) and the three Rabi frequencies for $|11\rangle$ are all integers up to a common factor $2\pi /t_{\text{g}}$; this leads to the following state evolution during $t\in[0,~t_{\text{g}}]$,
\begin{eqnarray}
 && |01\rangle\rightarrow e^{i\alpha\frac{t}{t_{\text{g}}}}\left(|v_+\rangle + e^{2i\pi  N\frac{t}{t_{\text{g}}}}|v_-\rangle \right)\rightarrow e^{i\alpha} |01\rangle,\label{interference01}\\
  && |11\rangle\rightarrow e^{i\beta\frac{t}{t_{\text{g}}}}\sum_{j=1}^3 e^{2i\pi  M_j \frac{t}{t_{\text{g}}}}|V_j\rangle \rightarrow e^{i\beta}|11\rangle, \label{interference}
  \end{eqnarray}
where $\alpha=-N\pi(1+\Delta/ \sqrt{\Omega^2+\Delta^2})$, $\beta=-N\pi(2\Delta+2V/3 )/ \sqrt{\Omega^2+\Delta^2}$, and $\{|v_\pm\rangle\}$ and $\{|V_1\rangle,~|V_2\rangle,~|V_3\rangle\}$ are eigenvectors of the Hamiltonians for the input states $|01\rangle$ and $|11\rangle$, respectively, as shown later. Here, $\Omega$, $\Delta$, and $V$ are the Rabi frequency, detuning, and van der Waals interaction~(vdWI) between Rydberg atoms, respectively. A physical picture similar to $|01\rangle$ applies to $|10\rangle$. Then, the population of the two-qubit system after interaction with laser radiation for a time of $t_{\text{g}}$ remains unchanged regardless of the initial state of the qubits, but phase shifts occur for certain initial states, as shown in Fig.~\ref{figure01}. An entangling gate emerges when the phase shifts to $\{|01\rangle,|10\rangle\}$ and $|11\rangle$ satisfy $2\alpha-\beta\neq 2j\pi$, where $j$ is an integer. As for the gate accuracy, there are two intrinsic fidelity errors in our method, namely, the Rydberg-state decay and a rotation error. The rotation error is below $10^{-5}$, thus can be neglected. Because there is no resonant transition to the Rydberg states, the population in Rydberg states is tiny. Consequently, the error caused by Rydberg-state decay can be easily reduced to the order of $10^{-5}$ under typical experimental conditions.  

With the help of single-qubit gates, $U_{1(2)}$ can be transformed to the C$_{\text{Z}}$ gate, which is a basic gate for universal quantum computing~\cite{Williams2011,Nielsen2000}. Our gate has the form of $U(\alpha,\beta)=\text{diag}\{1, e^{i\alpha}, e^{i\gamma}, e^{i\beta}\}$ written in the matrix form with the two-qubit basis $\{|00\rangle,|01\rangle,|10\rangle,|11\rangle \}$. The angles $\alpha$ and $\gamma$ are functions of the Rabi frequency $\Omega$ and detuning $\Delta$ of the Rydberg lasers, and $\beta$ is determined by $\Omega$, $\Delta$, and $V$. $\alpha$ is equal to $\gamma$ in $U_1$, and by varying $\Omega$, $\Delta$, and $V$, various combinations of $\alpha$ and $\beta$ can be achieved as shown later. When $|\alpha/2-\beta|$ is not too small, a combination of two $U_1(\alpha,\beta)$ and several single-qubit gates lead to a C$_{\text{Z}}$~\cite{PhysRevLett.89.247902}. For the $U_2$ gate, one can easily realize the cases where $|\beta-\alpha-\gamma\pm \pi/2|/\pi$ is an odd integer, so that two single-qubit phase gates and two $U_2$ form a C$_{\text{Z}}$ gate. 

The comparable magnitudes of $\Omega$, $\Delta$, and $V$ in our method allow a rapid implementation when both $\Omega$ and $V$ are near the feasible value of $2\pi\times10$~MHz~\cite{Gaetan2009}. We recognize that protocols of Rydberg quantum gates with single laser pulses had been proposed~\cite{Su2016,Su2017,Han2016}. Nevertheless, the method in \cite{Su2016,Su2017} is based on an effective theory when $V\gg \Omega$ and relies on pumping two atoms to Rydberg states with Rabi frequencies $\sim \Omega^2/V\ll \Omega$, thus not only lacking accuracy but also requiring long gate times accompanied by large Rydberg-state decay; as for the method in~\cite{Han2016}, although its speed is determined by $\Omega$ and can be large, the gate has a large rotation error in addition to the error caused by Rydberg-state decay. In contrast, our method is not only practical, but can easily attain a high fidelity that is necessary for scalable quantum computing.

The remainder of this article is structured as follows. In Sec.~\ref{sec02} and Sec.~\ref{sec03}, we outline the two protocols of $U_{1}$ and $U_{2}$, respectively, and show how a C$_{\text{Z}}$ gate is formed from them. In Sec.~\ref{sec04}, we look at the robustness of the interference method against the motion-induced dephasing of atomic transitions and variations of vdWI. In Sec.~\ref{sec05}, we discuss the prospect of extending our method to realize other types of quantum gates and using other types of Rydberg interactions. A summary is given in Sec.~\ref{sec06}.

\section{An entangling gate by a single laser pulse on both qubits}\label{sec02}
In this section, we show how quantum interference between different off-resonant Rabi oscillations can faithfully lead to the following quantum gate
\begin{eqnarray}
 U_1 &=& \left(
  \begin{array}{cccc}
    1& 0 & 0&0\\
    0 & e^{i\alpha} &0&0\\
    0 &0 & e^{i\alpha}&0\\
    0& 0 & 0&e^{i\beta}\\   
    \end{array} 
  \right) ,\label{gateU1}
  \end{eqnarray}
written in the ordered basis $\{|00\rangle,|01\rangle,|10\rangle,|11\rangle \}$. Several possible sets of values for $\alpha$ and $\beta$, determined by $\Omega$, $\Delta$, and $V$, can be easily found on a desktop computer, as shown later. By using the single-qubit phase gates
\begin{eqnarray}
  |1\rangle\rightarrow e^{-i\alpha}  |1\rangle
\end{eqnarray}
for both the control and target qubits, the gate in Eq.~(\ref{gateU1}) becomes diag$\{1,1,1, e^{i(\beta-2\alpha)}\}$, which is an entangling gate when $\beta-2\alpha$ is not a multiple of $2\pi$. 

The gate $U_1$ is implemented by applying a laser to the two qubits for the excitation $|1\rangle\rightarrow|r\rangle$, where $|r\rangle$ is a Rydberg state. After performing dipole approximation and rotating wave approximation in the rotation frame, the Hamiltonian becomes,
  \begin{eqnarray}
    \hat{H} &=&\Omega( |0r\rangle\langle01|+ |r0\rangle\langle10|+\text{H.c.})/2 \nonumber\\
    &&+\Delta(|0r\rangle\langle0r|+|r0\rangle\langle r0|  ) \nonumber\\
    &&+\hat{H}_{\text{v1}},\label{Hamil0110}
  \end{eqnarray}
  where $\hat{H}_{\text{v1}}$ is the Hamiltonian for the input state $|11\rangle$, given by
  \begin{eqnarray}
\hat{H}_{\text{v1}}&=&   \left(
  \begin{array}{ccc}
    V+2\Delta& \Omega/\sqrt2 & 0\\
    \Omega/\sqrt2 & \Delta&\Omega/\sqrt2\\
    0 &\Omega/\sqrt2 & 0
    \end{array}
  \right),
  \label{hamiltonian0}
\end{eqnarray}
in the basis of $|rr\rangle$, $(|1r\rangle+|r1\rangle)/\sqrt2$, and $|11\rangle$. We have assumed real laser Rabi frequencies in the equations above for simplicity; their phases become important when we study the Doppler dephasing later. For the widely used rubidium and cesium atoms, the qubit states $|0\rangle$ and $|1\rangle$ can be two hyperfine ground states with an energy separation of several gigahertz. The Rydberg state $|r\rangle$ can be a high-lying s- or d-orbital Rydberg state that is easily excited by two-photon excitation via a largely detuned low-lying p-orbital intermediate state. To avoid spontaneous emission from the intermediate state, a large gigahertz-scale detuning is necessary~\cite{Wilk2010,Isenhower2010,Zhang2010,Maller2015,Zeng2017}. Because the dipole moment between the low-lying intermediate state and $|r\rangle$ is quite small, the value of a practical $\Omega$ can not be very large. For this reason, we will assume that the two-photon Rydberg Rabi frequency $\Omega$ is smaller than $10\times2\pi$~MHz for a practical assessment of the achievable gate fidelity~\cite{Goerz2014,DeLeseleuc2018}; in the numerical analysis of the gate fidelity, we will be more conservative and use an $\Omega$ up to $0.8\times2\pi$~MHz that has been realized in quantum gates with Rydberg states of principal quantum numbers around 100~\cite{Zhang2010}.

Among the four input states, the state $|00\rangle$ remains unchanged because the qubit state $|0\rangle$ is not optically pumped. For the other input states, $\{|01\rangle,|10\rangle\}$ and $|11\rangle$ are governed by different Hamiltonians, and we will study their dynamics successively for each case.

\subsection{Phase accumulation of $\{|01\rangle,|10\rangle\}$ in response to detuned Rabi oscillations }\label{sec02a}
\begin{figure}
\includegraphics[width=3.2in]
{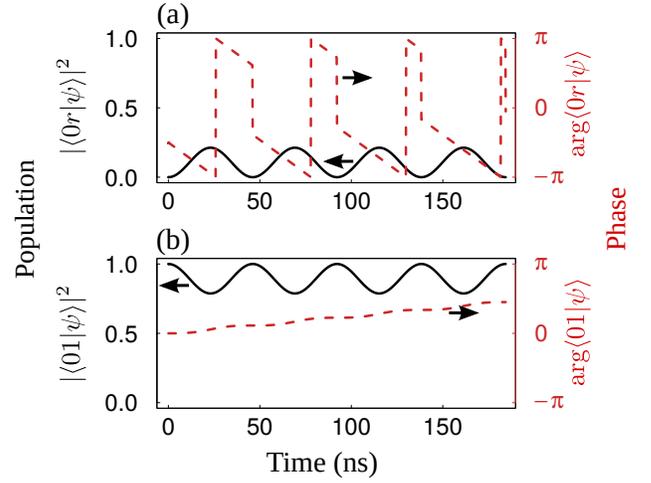}
\caption{Time evolution of the population and phase of the two components $|0r\rangle$~[in (a)] and $|01\rangle$~[in (b)] in the wavefunction $|\psi\rangle$ during the gate sequence of $U_1$ when the input state is $|01\rangle$. The solid~(dashed) curve denotes the population~(phase). Calculation is performed by using Eq.~(\ref{Hamil0110}) with the parameters in case 1 listed in Table~\ref{table1}. Here arg$(\cdot)$ gives the argument of a complex variable. The Hamiltonian evolves the state $|01\rangle$ back to itself exactly. Similar behavior exists for the input state $|10\rangle$. No population loss occurs for the input states $|01\rangle$ and $|10\rangle$ when the Rydberg-state decay and technical issues are ignored. \label{U1-01} }
\end{figure}

According to Eq.~(\ref{Hamil0110}), the time dynamics should be similar for the two input states $|01\rangle$ or $|10\rangle$. Taking $|01\rangle$ as an example, we cast the Hamiltonian for it into the form of a matrix,
\begin{eqnarray}
  \hat{H} &=&
  \left(\begin{array}{cc}
    \Delta& \Omega/2 \\
    \Omega/2 &0
    \end{array}
  \right)\label{IIAeq01}
\end{eqnarray}
in the basis $|0r\rangle$ and $|01\rangle$. The above Hamiltonian can be diagonalized with two Stark-shifted eigenstates, between which there is a transition frequency $\bar\Omega \equiv\sqrt{\Omega ^2+\Delta^2}$, as shown in Fig.~\ref{figure01new}. By choosing the pulse duration to be
\begin{eqnarray}
 t_{\text{g}}&=& 2N\pi/\bar\Omega , \label{gatetime01}
\end{eqnarray}
where $N$ is an integer, the input state $|\psi(0)\rangle=|01\rangle$ becomes~\cite{Shi2017}
\begin{eqnarray}
  |\psi(t_{\text{g}})\rangle &=&e^{i\alpha}|01 \rangle \label{trans01}
\end{eqnarray}
when the optical pumping completes, where
\begin{eqnarray}
 \alpha/\pi&=& -N(1+\Delta/ \bar\Omega).\label{alpha}
\end{eqnarray}
The above process is free from Rydberg blockade, thus its intrinsic accuracy is limited only by the fundamental Rydberg-state decay. Figure~\ref{U1-01} shows a numerical simulation of the dynamics for the input state $|01\rangle$ using the parameters listed in case 1 of Table~\ref{table1}. One finds that after four detuned Rabi cycles with period $2\pi/ \bar\Omega$, the state returns to itself with a phase accumulation shown in Eq.~(\ref{alpha}). $|10\rangle$ has a similar time evolution as the quantum interference patterns are the same for $|01\rangle$ and $|10\rangle$, shown in Fig.~\ref{figure01}.

\begin{figure}
\includegraphics[width=3.3in]
{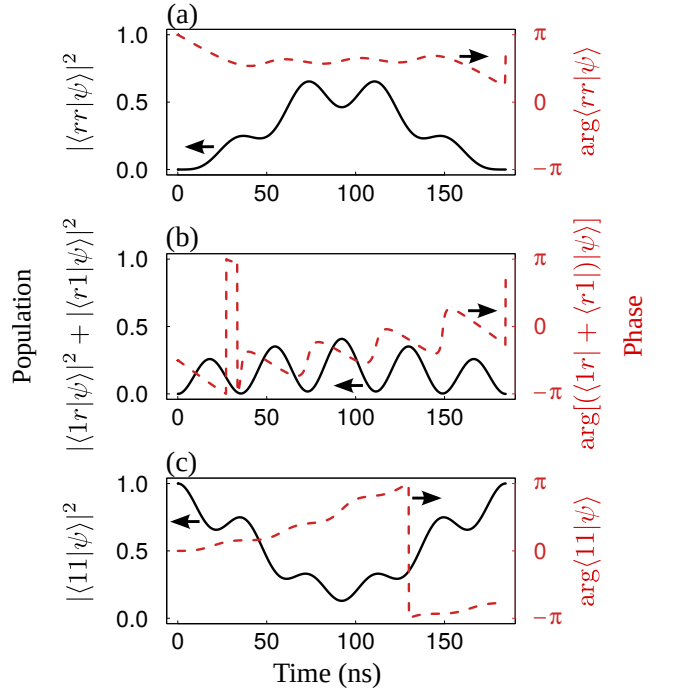}
\caption{The solid and dashed curves show the population and phase dynamics of the three components $|rr\rangle,(|1r\rangle+|r1\rangle)/\sqrt2 $, and $|11\rangle$ in (a), (b), and (c), respectively. This dynamics of the wavefunction $|\psi\rangle$ occurs during the gate sequence when the input state is $|11\rangle$ in the $U_1$ gate. The population loss is on the order of $10^{-10}$ at the end of the laser excitation. Parameters are the same with those in Fig.~\ref{U1-01}.  \label{U1-11} }
\end{figure}

\subsection{Phase accumulation of $|11\rangle$ in response to detuned Rabi oscillations }\label{sec02b}
For the input state $|11\rangle$, the Hamiltonian is given by $\hat{H}_{\text{v1}}$ in Eq.~(\ref{hamiltonian0}). Similar to Eq.~(\ref{IIAeq01}), $\hat{H}_{\text{v1}}$ can also be diagonalized as $\hat{H}_{\text{v1}} = \sum_{\chi=1,2,3}\epsilon_\chi |v_\chi \rangle\langle v_\chi |$. As shown in Sec.~\ref{sec02a}, it is the eigenvalues of the Hamiltonian that matter in the time evolution. By using the Shengjin equation for the cubic equation~\cite{Shengjin}, the eigenvalues of $\hat{H}_{\text{v1}}$ are given by
\begin{eqnarray}
  \epsilon_1 &=&\Delta+[V -2\mathscr{A}\cos\theta]/3,\nonumber\\
  \epsilon_{2(3)} &=&\Delta+[V+ 2\mathscr{A}\cos(\theta\pm\pi/3)] /3  ,
  \end{eqnarray}
where
\begin{eqnarray}
  \mathscr{A}&=& \sqrt{V^2+3(\Omega^2+\Delta^2+\Omega\Delta)},\nonumber\\
  \theta&=&\frac{1}{3} \arccos\frac{  \mathscr{B}}{2\mathscr{A}^3}
   ,\nonumber\\
  \mathscr{B} &=& 27\Omega^2(V/2+\Delta)+9(V+3\Delta)(2\Delta^2+V\Delta -\Omega^2)\nonumber\\
  &&-2(V+3\Delta)^3 .
\end{eqnarray}
To prevent distraction from the main theory, the eigenvectors of $\hat{H}_{\text{v1}}$ are not shown here. The initial state $|\psi(0)\rangle=|11\rangle$ can be written as
\begin{eqnarray}
  |\psi(0)\rangle &=& \sum_{\chi=1}^3\alpha_\chi|v_\chi \rangle,
  \end{eqnarray}
where $\alpha_\chi$ are functions $\Omega$, $\Delta$, and $V$. During the gate sequence, the above state becomes
\begin{eqnarray}
  |\psi(t)\rangle &=& \sum_{\chi=1}^3 \alpha_\chi e^{-it \epsilon_\chi}|v_\chi \rangle\nonumber\\
  &=& e^{-i(\Delta+V/3)t} \sum_{\chi=1}^3 \alpha_\chi e^{it \mathscr{C}_\chi} |v_\chi \rangle,
\end{eqnarray}
where
\begin{eqnarray}
  \mathscr{C}_1&=&2\mathscr{A}\cos\theta/3, \nonumber\\
  \mathscr{C}_2  &=&  -2\mathscr{A}\cos(\theta+\pi/3)/3, \nonumber\\
  \mathscr{C}_3  &=&  -2\mathscr{A}\cos(\theta-\pi/3)/3.
\end{eqnarray}
In order to cause the state to return to the ground state $|11\rangle$ after the laser pulse finishes, we consider the following condition
\begin{eqnarray}
 t_{\text{g}} \mathscr{C}_\chi&=&2M_\chi\pi, \label{gatetime02}
\end{eqnarray}
where $M_\chi$ are integers for $\chi=1,2,$ and 3. Then, when the laser excitation completes, the input state $|11\rangle$ becomes,
\begin{eqnarray}
  |\psi(t_{\text{g}})\rangle &=& e^{i\beta } |11\rangle,
\end{eqnarray}
where
\begin{eqnarray}
  \beta/\pi&=& -(\Delta+V/3) t_{\text{g}}/\pi,
\end{eqnarray}
and the latter can be expressed as
\begin{eqnarray}
  \beta/\pi&=&-N[2\Delta/\bar\Omega+2V/(3\bar\Omega) ] 
\end{eqnarray}
using Eq.~(\ref{gatetime01}).

 Similar to Fig.~\ref{U1-01}, Fig.~\ref{U1-11} shows a numerical simulation of the dynamics for the input state $|11\rangle$ with the parameters adopted from case 1 of Table~\ref{table1}. We find that the population leakage out of the computational basis is on the order of $10^{-10}$, which means that the rotation error is negligible. 

\subsection{Accurate and rapid entangling gates without individual addressing of atoms}\label{sec02c}
When the Rydberg blockade $V$, the duration $t_{\text{g}}$, the Rabi frequency $\Omega$, and the detuning $\Delta$ of the laser satisfy the conditions in Eqs.~(\ref{gatetime01}) and~(\ref{gatetime02}), a transformation of Eq.~(\ref{gateU1}) is realized. To minimize the gate error caused by decay of Rydberg state, a comparatively small gate time is preferred. For $V$, $\Omega$, and $\Delta$ on the order of $2\pi\times10$~MHz, a smaller $N$ leads to shorter gate times. Because it is difficult to analytically solve the set of equations in Eqs.~(\ref{gatetime01}) and~(\ref{gatetime02}) for a chosen set of $(N,M_1,M_2,M_3) $, we obtain a numerical solution as follows. We fix $\Omega/2\pi=10$~MHz, but vary $\Delta$ and $V$ when $t_{\text{g}}$ is given by Eq.~(\ref{gatetime01}), so that the transformation in Eq.~(\ref{trans01}) with $\alpha$ given in Eq.~(\ref{alpha}) is exact. An appropriate set of $\Delta$ and $V$ is first determined by the criteria of smaller rotation error, which is defined as
\begin{eqnarray}
E_{\text{ro}}&=&1/4-\left|\langle 11 |e^{-it_{\text{g}}\hat{H}_{\text{v1}}  } |11\rangle\right |^2/4.
\end{eqnarray}
To guarantee a large enough intrinsic gate fidelity, we not only discard all cases with $E_{\text{ro}}>10^{-7}$, but also choose cases with small enough Rydberg-state decay. An analytical approximation for the decay error~\cite{Zhang2012} is given by
\begin{eqnarray}
E_{\text{de}}&=&[T_{\text{r}}(01)+T_{\text{r}}(10) +T_{\text{r}}(11)+2 T_{\text{rr}}(11)]/(4\tau),
\end{eqnarray}
where $T_{\text{r}}(ab)$ is the time for the input state $|ab\rangle$ to be in a state like $|ar\rangle$ or $|rb\rangle$, while $T_{\text{rr}}(11)$ is the time for the input state $|11\rangle$ to be in $|rr\rangle$ during the gate sequence.

In Table~\ref{table1}, we list three sets of $V$, $\Omega$, and $\Delta$ where $E_{\text{ro}}<10^{-7}$ and $E_{\text{de}}<90$~ns$/\tau$, where $\tau$ is the lifetime of the Rydberg state $|r\rangle$. For completeness, we also list the gate times, the integers $N$ and $M_\chi$, $\chi=1-3$, and $\beta-2\alpha$. All cases in Table~\ref{table1} are entangling gates since the angle $\beta-2\alpha\neq 2\pi$. However, we note that, an exhaustive search has not been done to obtain Table~\ref{table1}, and in principle, many other cases exist with a high fidelity and a large gate speed. At a glance, it is strange to find a mismatch between the ordering of the three sets according to $t_{\text{g}}$ and that according to $E_{\text{de}}$ in Table~\ref{table1}. To understand this mismatch, we note that both the detuning $\Delta$ and the Rydberg blockade $V$ for case 2 are largest, so it has a very small population in the Rydberg state. This is why despite the longest gate duration for case 2, its decay error is smaller than that of case 3.

We can ignore the rotation error since it is smaller than $10^{-7}$ for the cases in Table~\ref{table1}. However, it is challenging to realize a very large $\Omega$ to ensure a smaller gate time. So, it is not practical to reduce the decay error beyond $10^{-6}$. Then, the intrinsic fidelity error is determined by the decay error only
\begin{eqnarray}
1-\mathcal{F}=E_{\text{de}}, \label{fidelityError01}
\end{eqnarray}
which is equal to $45.5$~ns$/\tau$ for case 1 in Table~\ref{table1}. For $s$ or $d$ orbital Rydberg states with principal quantum numbers around 100, their lifetimes are about 1 millisecond in a temperature of 4.2 K~\cite{Beterov2009}. This means that the error of the gate fidelity can reach $4\times10^{-5}$ for case 1. We note that even if it is difficult to have a Rabi frequency of $2\pi\times10$~MHz, a reduction by a factor of 10 of the three parameters $V$, $\Omega$, and $\Delta$~(or only the parameter $\tau$) will increase the decay error by a factor of $10$. This still guarantees a gate fidelity of about $0.9991-0.9995$ for all three cases in Table~\ref{table1}.

In the analysis above, we have neglected the leakage of population to other Rydberg states near $|r\rangle$. This is because as shown in~\cite{Shi2017}, the leakage error can be made negligible by varying $\Omega$ slightly from the chosen value~(when $\Delta$ and $V$ are changed with the same ratio for our gate). The reason is that the leakage of population to a detuned energy level is given by $[x\sin(y/x)]^2$, where $y$ is the area of the pulse, $x=\Omega/\sqrt{\Omega^2+\delta^2}$, and $\delta$ is the detuning of a nearby Rydberg state~\cite{Shi2018prapp2}. This method can effectively remove such population leakage especially because nearby Rydberg states usually appear almost symmetrically in pairs around $|r\rangle$. Another method to reduce such leakage error is by shifting away the nearby levels via external fields~\cite{Petrosyan2017}. By either of these methods, the population leakage to nearby levels can be suppressed and Eq.~(\ref{fidelityError01}) gives the intrinsic fidelity error for our entangling gates.

\subsection{Construct a C$_{\text{Z}}$ from $U_1$ and single-qubit gates}\label{sec02d}
For neutral atoms, it is a well-established technique to transform a C$_{\text{Z}}$ to a CNOT~\cite{Isenhower2010}, which is the most well-known entangling gate for forming a universal set of quantum gates with single-qubit gates~\cite{Williams2011,Nielsen2000}. For this reason, we show how a C$_{\text{Z}}$ can be constructed from $U_1$ and single-qubit gates below, where $U_1$ is given by Eq.~(\ref{gateU1}). Because different angles $\beta-2\alpha$ in the $U_1$ gate require different numbers of $U_1$, we choose the $U_1$ with parameter set 1 of Table~\ref{table1} as an example.

We follow the method of Ref.~\cite{PhysRevLett.89.247902} to show how a C$_{\text{Z}}$ is obtained from $U_1$. For brevity, we define
\begin{eqnarray}
  \hat{\mathscr{P}}_\phi &=& \left( \begin{array}{cc}
    1& 0 \\
    0 & e^{i\phi}
    \end{array} \right) \label{phasegate}
\end{eqnarray}
as a phase shift gate, and $\hat{\mathscr{R}}_{\mathbf{n}}(\theta) \equiv e^{-i\theta\mathbf{n}\cdot(\mathbf{x}\sigma_1 +\mathbf{y}\sigma_2+ \mathbf{z}\sigma_3 )/2 }$ as a rotation of angle $\theta$ around the axis $\mathbf{n}$, where $\sigma_{j}$ is the Pauli matrix with $j=1-3$. We further define the following phase change on the qubits:
\begin{eqnarray}
\hat{O}_1&=&[  \hat{\mathscr{P}}_{-\beta/2}]_{\text{c}}\otimes[ \hat{\mathscr{P}} _{-\alpha} ]_{\text{t}},\nonumber\\
\hat{O}_2&=&[ \hat{\mathscr{P}}_{-\beta}]_{\text{c}}\otimes[  \hat{\mathscr{P}}_{-2\alpha} ]_{\text{t}},\label{O1definition}
\end{eqnarray}
and construct a controlled-$\pi$ rotation about $\mathbf{z}$ by using the following two gates:
\begin{eqnarray}
\hat{O}_1 U_1&=&|0\rangle\langle0|\otimes \hat{I} + |1\rangle\langle1|\otimes \hat{\mathscr{R}}_{\mathbf{z}}(\beta-2\alpha),\nonumber\\
\hat{O}_2 U_1^2&=&|0\rangle\langle0|\otimes \hat{I} + |1\rangle\langle1|\otimes\hat{\mathscr{R}}_{\mathbf{z}}(2\beta-4\alpha) .\label{sec02D_EQ01}
\end{eqnarray}
Here the subscript c~(t) denotes the control~(target) qubit, and the operator left~(right) of ``$\otimes$'' acts on the control~(target) qubit. The reason for constructing $\hat{O}_2 U_1^2$ is as follows. According to Ref.~\cite{PhysRevLett.89.247902}, it is necessary to distinguish whether $|\alpha-\beta/2|$ is larger than $\pi/6$; if not, at least four $U_1$ gates are needed. A special case of $|\alpha-\beta/2|=2\pi/3$ has been studied in~\cite{PhysRevLett.89.247902}. Because $|\alpha-\beta/2|<\pi/6$ for the parameter set 1 in Table~\ref{table1}, a new rotation with twice of the angle $|\alpha-\beta/2|$ that is larger than $\pi/6$ is needed. For this reason, we should operate two $U_1$ so that we get $U_1^2$; In our interference method, this can be done by doubling the pulse duration so that the angles $\alpha$ and $\beta$ in Eq.~(\ref{gateU1}) are doubled.

Five steps can form a C$_{\text{Z}}$ starting from the two controlled-rotations in Eq.~(\ref{sec02D_EQ01}). To do this, we need two single-qubit gates $\hat{A}$ and $\hat{B}$ for the target qubit and a phase gate $\hat{\mathscr{P}}_{\pi/2}$ for the control. (i) First, apply $\hat{A}$ to rotate the axis of the rotation operator $\hat{O}_1 U_1$ from $\mathbf{z}$ to $\mathbf{n}_{2}$. (ii) Next, combine $\hat{O}_1 U_1$ with the gate from the first step to get a controlled rotation of angle $\pi+2(2\alpha-\beta)]$ about an axis $\mathbf{n}_{12}$. (iii) Apply $\hat{B}$ to rotate the axis of the rotation operator of the second step from $\mathbf{n}_{12}$ to $\mathbf{z}$. (iv) Combine the rotation obtained in the last step and $\hat{O}_2 U_1^2$. (v) Apply the phase gate $\hat{\mathscr{P}}_{\pi/2}$ for the control qubit. These steps can be represented mathematically by
\begin{eqnarray}
  \text{C}_{\text{Z}} &   = & [\hat{\mathscr{P}}_{\pi/2}\otimes\hat{I}]  \hat{D}(\hat{O}_2 U_1^2 ),  \label{U1toCZ}
\end{eqnarray}
where $\hat{D}$ represents steps (i), (ii), and (iii), and is given by
\begin{eqnarray}
  \hat{D} &=&\left\{  [\hat{I} \otimes \hat{B}][\hat{O}_1U_1 (\hat{I} \otimes \hat{A})\hat{O}_1 U_1  (\hat{I} \otimes \hat{A}^\dag) ][\hat{I} \otimes \hat{B}^\dag]\right\}. \nonumber
\end{eqnarray}  
As shown in Appendix~\ref{appA}, one can solve for $\hat{A} =\hat{\mathscr{R}}_{\mathbf{y}}(\theta_1)  $ and $ \hat{B} =\hat{\mathscr{R}}_{\mathbf{z}}(\theta_3) \hat{\mathscr{R}}_{\mathbf{y}}(\theta_2)    $, where
    \begin{eqnarray}
      \theta_1 &=& \arccos  a,\nonumber\\
      \theta_2 &=& \arccos \frac{\sin\left(\beta-2\alpha\right) (1+ a)  }{2\cos\left(2\alpha-\beta\right)},\nonumber\\
      \theta_3&=& \arctan\frac{2 \sin^2\left(\frac{\beta}{2}-\alpha\right)   }{\sin\left(\beta-2\alpha\right)} ,  \nonumber\\
      a &=& \frac{ \cos^2\left(\frac{\beta}{2}-\alpha\right) + \sin\left(2\alpha-\beta\right)}{\sin^2\left(\frac{\beta}{2}-\alpha\right)} .
    \end{eqnarray}
    The angle $\theta_1$ is a function of $\beta/2-\alpha$, and the above solution applies for $\beta/2-\alpha>\pi/8$. Details of the derivation for the above angles and the expressions for $\mathbf{n}_{2}$ and $\mathbf{n}_{12}$ are given in Appendix~\ref{appA}.

The above method demonstrates the basic procedure for constructing a C$_{\text{Z}}$ from $U_1$ and single-qubit gates, and can be used for any $U_1$ gate with a specific $\beta-2\alpha$. For instance, to construct a C$_{\text{Z}}$ from $\hat{O}_1U_1$ of case 2 in Table~\ref{table1}, we only need two $\hat{O}_1U_1$ because $|\beta-2\alpha|>\pi/2$. The following four steps achieve this: use a single-qubit gate $\hat{A}'$ for the target to rotate the axis of the rotation $\hat{O}_1U_1$ from $\mathbf{z}$ to $\mathbf{n}_{2}'$, combine it with another $\hat{O}_1U_1$ to form a rotation of angle $\pi$ around the axis $\mathbf{n}_{12}'$, then use $\hat{B}$ to rotate the axis from $\mathbf{n}_{12}'$ back to $\mathbf{z}$, and finally use a phase shift gate for the control qubit to reach a C$_{\text{Z}}$. Similar methods shown in Appendix~\ref{appA} can be used to find $\hat{A}'$, $\hat{B}'$, $\mathbf{n}_{2}'$ and $\mathbf{n}_{12}'$.


\section{An entangling gate by one laser pulse on each of the two qubits}\label{sec03}
In this section, we show how quantum interference between different off-resonant Rabi oscillations can lead to the following quantum gate
\begin{eqnarray}
 U_{2}&=& \left(
  \begin{array}{cccc}
    1& 0 & 0&0\\
    0 & e^{i\alpha} &0&0\\
    0 &0 & e^{i\gamma}&0\\
    0& 0 & 0&e^{i\beta}\\   
    \end{array} 
  \right) \label{gateU2}
  \end{eqnarray}
written in the ordered basis $\{|00\rangle,|01\rangle,|10\rangle,|11\rangle \}$. By using single-qubit phase gates,
\begin{eqnarray}
  |1\rangle_{\text{c}}\rightarrow e^{-i\gamma}  |1\rangle_{\text{c}},~~|1\rangle_{\text{t}}\rightarrow e^{-i\alpha}  |1\rangle_{\text{t}},
\end{eqnarray}
the gate in Eq.~(\ref{gateU2}) becomes $U_{2}'=$diag$\{1,1,1, e^{i(\beta-\alpha-\gamma)}\}$. The angle $\beta-\alpha-\gamma$ can be tuned to a desired value in $(0,2\pi)$ by adjusting the laser parameters and the Rydberg blockade used in the protocol. We would like to obtain a high enough fidelity for the process, but this would impose strict gate times and achievable values of $\beta-\alpha-\gamma$. By a simple numerical search, we find that $\beta-\alpha-\gamma$ can be tuned with a high accuracy to $\pm\pi/2$, then, two $U_2'$ can be used to construct a C$_{\text{Z}}$. Compared to $U_1$, this method of realizing the C$_{\text{Z}}$ is relatively simple, but it comes with the complexity of single-site qubit addressing.

The gate is implemented by applying on each of the two qubits one laser pulse for the excitation $|1\rangle\rightarrow|r\rangle$. The pulse duration $t_k$, Rabi frequency $\Omega_k$, and detuning $\Delta_k$ of the laser for the qubit $k$ can be adjusted, where $k=$ c or t. After performing dipole approximation and rotating wave approximation in the rotating frame, the Hamiltonian reads,
  \begin{eqnarray}
    \hat{H} &=& \hat{H}_{\text{c}} + \hat{H}_{\text{t}}+\hat{H}_{\text{v}} .\label{Hamil_U2}
  \end{eqnarray}
  Here ,
  \begin{eqnarray}
     \hat{H}_{\text{c}} &=&\left\{\begin{array}{ll}
    \frac{\Omega_{\text{c}}}{2}(|r0\rangle\langle10|+\text{H.c.}) +\Delta_{\text{c}}|r0\rangle\langle r0|  ) , &\text{for } t\in[0,t_{\text{c}}],\\
    0, &\text{otherwise},
    \end{array}
    \right. \nonumber\\
     \hat{H}_{\text{t}} &=&\left\{\begin{array}{ll}
    \frac{\Omega_{\text{t}}}{2}(|0r\rangle\langle01|+\text{H.c.}) +\Delta_{\text{t}}|0r\rangle\langle0r|  ) , &\text{for } t\in[0,t_{\text{t}}],\\
    0, &\text{otherwise}.
    \end{array}
    \right. \nonumber\\ \label{Hct0110}
  \end{eqnarray}
 We define Min$(x,y)$ as a function that returns the smaller of $x$ and $y$. For $t\in[0,$Min$(t_{\text{c}},~t_{\text{t}})]$,  $\hat{H}_{\text{v}}$ is given by
 \begin{eqnarray}
     \hat{H}_{\text{v2}} &=& \left(
  \begin{array}{cccc}
    V+\Delta_{\text{c}}+\Delta_{\text{t}}& \Omega_{\text{t}}/2 &  \Omega_{\text{c}}/2 & 0\\
\Omega_{\text{t}}/2 &\Delta_{\text{c}}& 0&  \Omega_{\text{c}}/2 \\
\Omega_{\text{c}}/2  &0&\Delta_{\text{t}}&  \Omega_{\text{t}}/2 \\
    0& \Omega_{\text{c}}/2 &  \Omega_{\text{t}}/2 & 0
    \end{array}
  \right), \label{U2Hv}
  \end{eqnarray}
written in the basis of $\{|rr\rangle, |r1\rangle, |1r\rangle, |11\rangle\}$. However, $\hat{H}_{\text{v}}$ becomes
  \begin{eqnarray}
   \hat{H}_{\text{vt}} &=& \left(
  \begin{array}{cccc}
    V+\Delta_{\text{t}}& \Omega_{\text{t}}/2 & 0& 0\\
\Omega_{\text{t}}/2 &0& 0&  0\\
0  &0&\Delta_{\text{t}}&  \Omega_{\text{t}}/2 \\
    0&0 &  \Omega_{\text{t}}/2 & 0
    \end{array}
  \right) \label{U2Hv-t}
  \end{eqnarray}
  for the duration $[t_{\text{c}}, t_{\text{t}}]$ if $t_{\text{c}}<t_{\text{t}}$, and
  \begin{eqnarray}
   \hat{H}_{\text{vc}} &=& \left(
  \begin{array}{cccc}
       V+\Delta_{\text{c}}& 0 &  \Omega_{\text{c}}/2 & 0\\
0&\Delta_{\text{c}}& 0&  \Omega_{\text{c}}/2 \\
\Omega_{\text{c}}/2  &0&0& 0 \\
    0& \Omega_{\text{c}}/2 &  0& 0
    \end{array}
  \right) \label{U2Hv-c}
  \end{eqnarray}
  for the duration $[t_{\text{t}}, t_{\text{c}}]$ if $t_{\text{t}}<t_{\text{c}}$.

Because the two cases in Eqs.~(\ref{U2Hv-t}) and~(\ref{U2Hv-c}) are equivalent, we consider the case in Eq.~(\ref{U2Hv-t}) as an example. Furthermore, we impose the following condition: when the laser excitation on the control qubit completes, there is no population in either $|rr\rangle$ or $|r1\rangle$ so that Eq.~(\ref{U2Hv-t}) becomes
  \begin{eqnarray}
  \hat{H}_{\text{vt}} &=& \left(
  \begin{array}{cc}
       \Delta_{\text{t}}&   \Omega_{\text{t}}/2 \\
\Omega_{\text{t}}/2  & 0 
    \end{array}
  \right) \label{U2Hv-c2}
  \end{eqnarray}
in the basis of $\{ |1r\rangle,  |11\rangle\}$.

\subsection{Phase accumulation of $\{|01\rangle,|10\rangle\}$ }\label{sec03a}
Following the study in Sec.~\ref{sec02a}, the pulse durations for the control and target are chosen as
\begin{eqnarray}
 t_{\text{c}}&=& 2N_{\text{c}}\pi/\bar\Omega_{\text{c}} , \nonumber\\
 t_{\text{t}}&=& 2N_{\text{t}}\pi/\bar\Omega_{\text{t}} , \label{gatetimeU2}
\end{eqnarray}
where $N_{\text{c}}$ and $N_{\text{t}}$ are integers, and
\begin{eqnarray}
\bar\Omega_{\text{c}} &=& \sqrt{\Omega_{\text{c}} ^2+\Delta_{\text{c}}^2},\nonumber\\
\bar\Omega_{\text{t}} &=& \sqrt{\Omega_{\text{t}} ^2+\Delta_{\text{t}}^2}.\nonumber
\end{eqnarray}
Similar to the derivation of Eq.~(\ref{alpha}), one finds that the angles $\alpha$ and $\gamma$ in Eq.~(\ref{gateU2}) are given by
\begin{eqnarray}
  \alpha/\pi&=& -N_{\text{c}}(1+\Delta_{\text{c}}/ \bar\Omega_{\text{c}}), \nonumber\\
  \gamma/\pi&=& -N_{\text{t}}(1+\Delta_{\text{t}}/ \bar\Omega_{\text{t}}).  \label{alpha-U2}
\end{eqnarray}

\subsection{Phase accumulation of $|11\rangle$ }\label{sec03b}
For $t_{\text{c}}<t_{\text{t}}$, the time evolution of $|11\rangle$ is governed by Eq.~(\ref{U2Hv}) when $t\in(0,~t_{\text{c}}]$, and by Eq.~(\ref{U2Hv-c2}) when $t\in(t_{\text{c}},~t_{\text{t}}]$ if there is no population in either $|rr\rangle$ or $|r1\rangle$ upon the completion of the laser excitation on the control qubit. This latter condition can be written as
\begin{eqnarray}
 0&=& \left|\langle rr| e^{-it_{\text{c}}\hat{H}_{\text{v2}}  } |11\rangle\right|^2, \nonumber\\
 0&=& \left|\langle r1| e^{-it_{\text{c}}\hat{H}_{\text{v2}}  } |11\rangle\right|^2. \label{U2-cond2}
\end{eqnarray}
At the end of the laser excitation for the target qubit, it is also necessary to have the following condition
\begin{eqnarray}
 1&=&  \left|\langle 11|   e^{-i(t_{\text{t}}-t_{\text{c}})\hat{H}_{\text{vt}}  } e^{-it_{\text{c}}\hat{H}_{\text{v2}}  }  |11\rangle\right|^2. \label{U2-cond3}
\end{eqnarray}
Finally, in order to reach the phase accumulation $\beta-\alpha-\gamma=\pm\pi/2$ with $\beta$ given by
\begin{eqnarray}
 \beta&=&\text{arg}\left[\langle 11|   e^{-i(t_{\text{t}}-t_{\text{c}})\hat{H}_{\text{vt}}  } e^{-it_{\text{c}}\hat{H}_{\text{v2}}  }  |11\rangle\right],\nonumber
\end{eqnarray}
where arg$(\cdot)$ returns the argument of a complex variable, we need
\begin{eqnarray}
\beta&=&\pm\pi/2 +\alpha+\gamma.\label{U2-cond4}
\end{eqnarray}

\begin{figure}
\includegraphics[width=3.2in]
{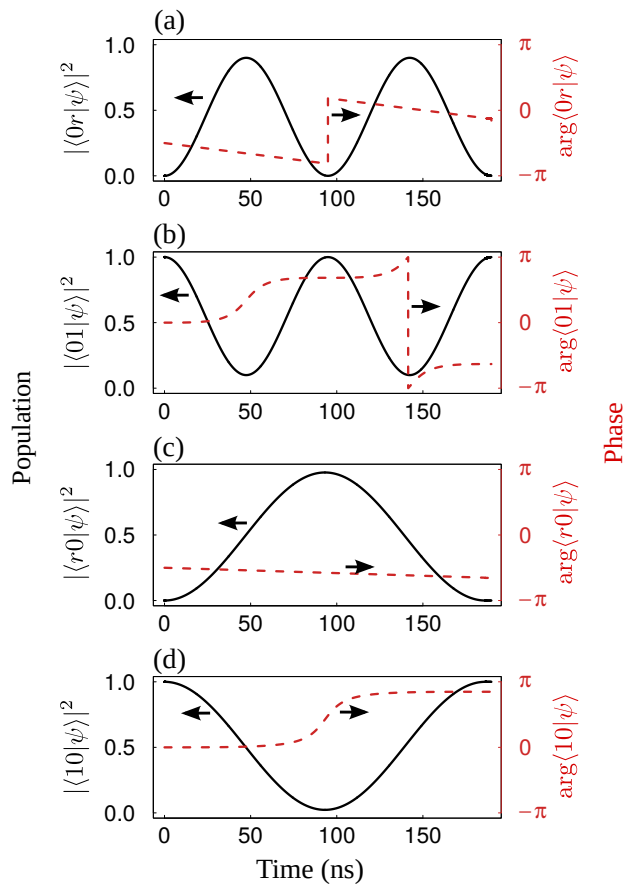}
\caption{(a) and (b) show the time evolution of the population~(by solid curves) and phase~(by dashed curves) of the components $|0r\rangle$ and $|01\rangle$, respectively, when the input state is $|01\rangle$ during the gate sequence of $U_2$. (c) and (d) show the time evolution of the population and phase of the components $|r0\rangle$ and $|10\rangle$, respectively, when the input state is $|10\rangle$. Calculation is performed by using the parameters in case 1 listed in Table~\ref{table2} with the phases of $\Omega_{\text{c}}$ and $\Omega_{\text{t}}$ equal to $\pi$ and $0$, respectively~(their phases can be arbitrary). No population loss occurs for the input states $|01\rangle$ and $|10\rangle$ when the Rydberg-state decay and technical issues are ignored. \label{U2-0110} }
\end{figure}

\begin{figure}
\includegraphics[width=3.3in]
{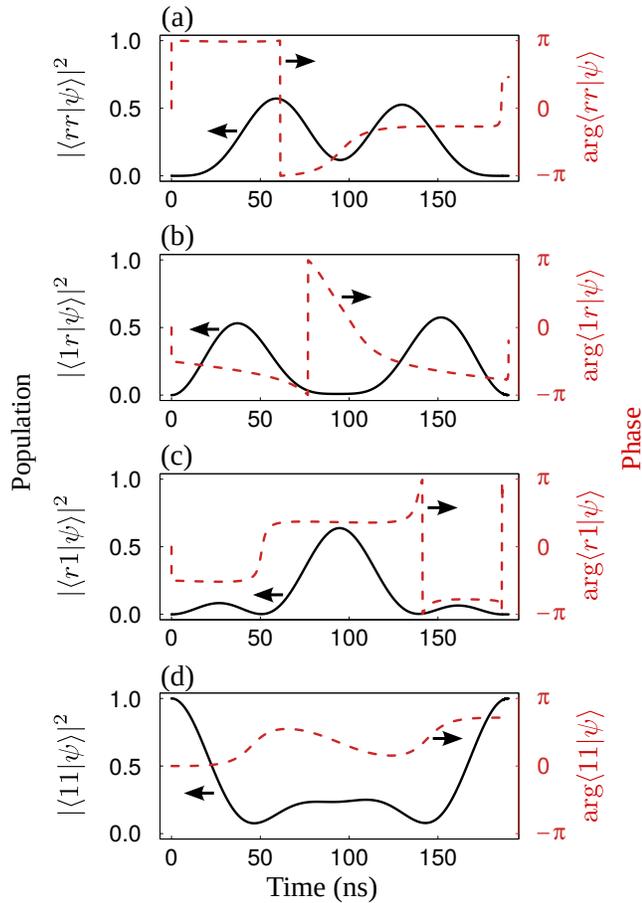}
\caption{(a), (b), (c), and (d) show the time evolution of the population~(by solid curves) and phase~(by dashed curves) of the four components $|rr\rangle,|1r\rangle,|r1\rangle $, and $|11\rangle$ in the wavefunction $|\psi\rangle$ during the gate sequence when the input state is $|11\rangle$ for the $U_2$ gate. Calculation is performed by using the same set of parameters as used in Fig.~\ref{U2-0110}. The population loss is on the order of $10^{-6}$ at the end of the laser excitation.  \label{U2-11} }
\end{figure}
\begin{table*}[ht]
  \centering
  \begin{tabular}{lccccc|ccccc}
    \hline
    \hline   \text{Case}&     $ \frac{|\Omega_{\text{c}}|}{2\pi}$~(MHz)&  $\frac{ \Delta_{\text{c}}}{2\pi}$ (MHz)  &     $ \frac{|\Omega_{\text{t}}|}{2\pi}$~(MHz)&  $\frac{ \Delta_{\text{t}}}{2\pi}$ (MHz)  &$\frac{  V}{2\pi} $ (MHz) &  $\frac{\beta-\alpha-\gamma}{\pi} $& $ (N_{\text{c}},~N_{\text{t}} ) $  &$(t_{\text{c}},~t_{\text{t}})$ (ns) &  $E_{\text{ro}}$& $E_{\text{de}}$ \\ \hline
      1&5.306482	 & 0.8152206 & 10  & 3.329994 & -5.442221 &4.5 & (1,~2)  &(186,~ 190)&$3.80\times10^{-6}$ &  86.8ns/$\tau$ 	 \\
      2&5.306482	 & -0.8152206 & 10  & -3.329994 & 5.442221 &1.5 & (1,~2)  &(186,~ 190)&$3.80\times10^{-6}$ &  86.8ns/$\tau$ 	 \\
      3& 3.331812  &  0.7475813&   10 &  1.825131& -3.418967 &4.5 & (1,~3)  &(293,~ 295)&$3.42\times10^{-8}$ &  140ns/$\tau$ 	 
      \\ 
      4& 3.331812  &  -0.7475813&   10 &  -1.825131& 3.418967 &3.5 & (1,~3)  &(293,~ 295)&$3.42\times10^{-8}$ &  140ns/$\tau$ 	 
      \\ \hline \hline
  \end{tabular}
  \caption{  \label{table2} Two sets of parameters $|\Omega_{\text{c}}|, \Delta_{\text{c}},| \Omega_{\text{t}}|, \Delta_{\text{t}}, V$, and $N_{\text{c}},N_{\text{t}}$ for realizing $U_2$ that satisfy Eq.~(\ref{alpha-U2}) exactly, and simultaneously satisfy Eqs.~(\ref{U2-cond2}),~(\ref{U2-cond3}), and~(\ref{U2-cond4}) approximately with a rotation error small than $10^{-5}$. A simultaneous scaling up or down of $|\Omega_{\text{c}}|, \Delta_{\text{c}}, |\Omega_{\text{t}}|, \Delta_{\text{t}}$, and $ V$ can still give rise to the same gate. Note that the phases of $\Omega_{\text{c}}$ and $\Omega_{\text{t}}$ can be arbitrary. }
  \end{table*}

\subsection{Realize $U_2$ with individual qubit addressing} \label{sec03c}
For any given $(\alpha,\gamma,\beta)$, $U_2$ is achievable as long as there are solutions, $(\Omega_{\text{c}},\Delta_{\text{c}},\Omega_{\text{t}},\Delta_{\text{t}},V, N_{\text{c}},N_{\text{t}})$, to the six equations in Eqs.~(\ref{alpha-U2}),~(\ref{U2-cond2}),~(\ref{U2-cond3}), and~(\ref{U2-cond4}). Given the fact that there are six equations with seven variables, various solutions exist for a pure mathematical argument. However, if the gate time is too large, sizable Rydberg-state decay arises which reduces the fidelity of the gate. Thus, we would like to find solutions for short enough gate times where the rotation error is minimal. To do this, a numerical search can be employed as described in Sec.~\ref{sec02c}. 

To quantify the rotation error of the gate fidelity in the computer search, we use the following definition of the rotation error for numerical calculation~\cite{Pedersen2007},
\begin{eqnarray}
 E_{\text{ro}} &=& 1-\frac{1}{20}\left[  |\text{Tr}(U_2^\dag \mathscr{U}_2)|^2 + \text{Tr}(U_2^\dag \mathscr{U}_2\mathscr{U}_2^\dag U_2 ) \right], \label{fidelityErrorPederson2007}
\end{eqnarray}
where $\mathscr{U}_2$ is the actual gate in the matrix form shown in Eq.~(\ref{gateU2}), and Tr denotes the trace of a matrix. Because the numerical search starts from setting the gate times in Eq.~(\ref{gatetimeU2}) with the corresponding angles $\alpha$ and $\gamma$ in Eq.~(\ref{alpha-U2}), $\mathscr{U}_2$ has the form of diag$\{1, e^{i\alpha}, e^{i\gamma}, \epsilon e^{i\beta'}\}$, where $\epsilon$ can be smaller than the desired value of $1$ and $\beta'$ can deviate from $\beta$. Then, the intrinsic rotation error is given by
\begin{eqnarray}
E_{\text{ro}} &=&1- \frac{1}{20}\left[  \left|3+ \epsilon e^{i\beta'-i\beta}\right|^2 + 3 + \epsilon^2\right]. \nonumber
\end{eqnarray}
Note that for the study of $U_1$ in Sec.~\ref{sec02}, the intrinsic rotation error arises solely from the population leakage because there is no error in the angles $(\alpha, \beta)$.

Several hours of search performed on a desktop computer resulted in four sets of parameters, shown in Table~\ref{table2}. In the numerical search, we restrict the Rabi frequencies to be less or equal to $10\times2\pi$~MHz. Cases with rotation errors above $10^{-5}$ and with decay errors over $150$ns$/\tau$ were discarded. The detunings $(\Delta_{\text{c}},\Delta_{\text{t}})$ and vdWI $V$ for the two cases 1 and 2 (or 3 and 4) in Table~\ref{table2} are opposite to each other, and the only observable difference between the two quantum gates is in the angle $\beta-\alpha-\gamma$. At a glance, these two cases in Table~\ref{table2}  are equivalent to each other since both of them result in an angle $\beta-\alpha-\gamma=\pm\pi/2$. The reason to list both of them is that for a chosen Rydberg state $|r\rangle$, $V$ is either positive or negative. Then, the two pairs for cases (1,2) and (3,4) in Table~\ref{table2} reveal that for any chosen Rydberg state, it is possible to achieve the $U_2$ gate with a high accuracy.  

As an example, we take the parameters in case 1 of Table~\ref{table2} to demonstrate how the gate works. The time evolution of the three input states $|01\rangle,|10\rangle$, and $|11\rangle $ is given in Figs.~\ref{U2-0110} and~\ref{U2-11}. Although we have assumed that the two external pulses on the control and target qubits begin at exactly the same moment, we found that the gate accuracy does not suffer if it is not strictly satisfied. 

Again, we emphasize that the listed cases in Table~\ref{table2} may not be the only cases to realize $U_2$. If we allow larger rotation and decay errors, many more cases can be found with gate errors on the order of $10^{-3}-10^{-4}$ with experimentally feasible conditions. Furthermore, it should be possible to find more high-fidelity gates $U_2$ in a more exhaustive numerical search. Here we have shown the basic method to realize the interference quantum gate based on Rydberg blockade, upon which experimentalists can find a set of parameters in their own setup.

\subsection{Construct a C$_{\text{Z}}$ from $U_2$ and single-qubit gates}\label{sec03d}
The method for constructing a C$_{\text{Z}}$ from $U_2$ is much simpler compared with that from $U_1$. This can be done by using two $U_2$ when assisted by two phase gates defined in Eq.~(\ref{phasegate}). For each $U_2$ in Table~\ref{table2}, a C$_{\text{Z}}$ is realized by applying a phase change operation on two subsequent $U_2$ gates,
\begin{eqnarray}
 \text{C}_{\text{Z}} &=&[  \hat{\mathscr{P}}_{-2\gamma}]_{\text{c}}\otimes[  \hat{\mathscr{P}}_{-2\alpha}]_{\text{t}}U_2^2.
\end{eqnarray}
If a CNOT is desired, two extra rotations of $\pm\pi/2$ about $\mathbf{y}$ axis in the target can be used. 

In comparison to the relatively complex method of realizing  a C$_{\text{Z}}$ by using $U_1$ shown in Sec.~\ref{sec02d}, the $U_2$ gate seems more favorable because it needs much fewer single-qubit gates. However, the realization of $U_2$ depends on single-site addressability. So the $U_1$ gate may be more useful because it is easy to realize high-fidelity single-qubit gates in arrays of neutral atom~\cite{Xia2015,Wang2016}.

\begin{figure}
\includegraphics[width=2.8in]
{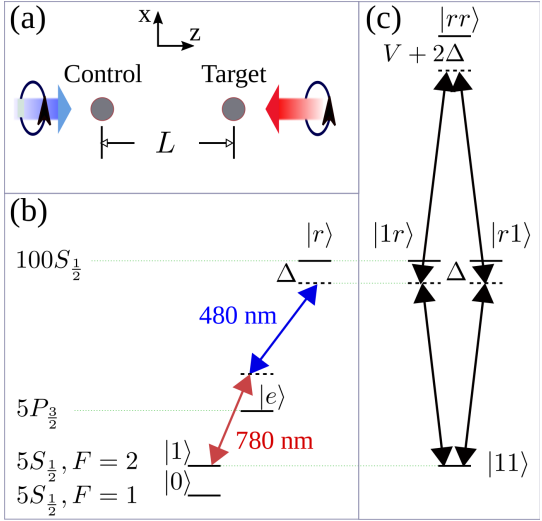}
 \caption{Scheme for realizing $U_1$ by circularly polarized lasers.  (a) Geometry of the setup, (b) scheme of laser excitation, and (c) state excitation for the input state $|11\rangle$ in the gate protocol for $U_1$.    \label{U1-configuration} }
\end{figure}

\begin{figure}
\includegraphics[width=3.3in]
{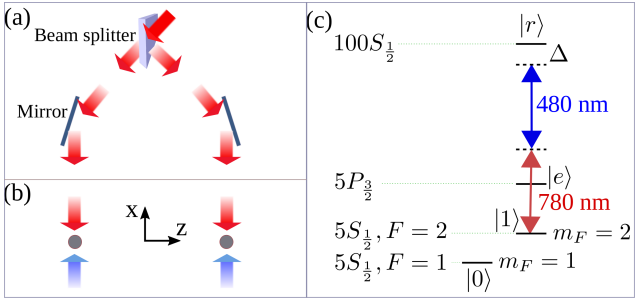}
 \caption{Scheme for realizing $U_1$ or $U_2$ by linearly polarized~(along $\mathbf{z}$) lasers. (a) Method to split one laser beams to two of equal strength for the realization of $U_1$, (b) geometry of the setup for the atomic qubits, and (c) scheme of laser excitation.    \label{U1-configuration-2ed} }
\end{figure}
\section{Resilience to motion-induced dephasing of atomic transitions}\label{sec04}
The main advantage of the interference method is that there is only one pulse of laser excitation for each qubit, so that no gap time is allowed in which the qubit is left in the Rydberg states. This is why the Doppler dephasing is minimized~\cite{DeLeseleuc2018}. To verify this assertion, we numerically investigate the error of our gate protocols by considering, for example, a qubit system of two $^{87}$Rb atoms and the Rydberg state $|r\rangle=|100s_{1/2},m_J=1/2,m_I=3/2\rangle$. This Rydberg state is relatively high, but the excitation of an even-parity Rydberg state of principal quantum numbers around $100$ has been achieved in, e.g., Refs.~\cite{Isenhower2010,Dudin2012,Zhang2010}.

A proper investigation should be based on a practical atomic and optical configuration. In order to incur a less severe Doppler dephasing, we consider the excitation of Rydberg states by two counter-propagating laser fields~\cite{Saffman2016}, shown in Fig.~\ref{U1-configuration}(a) and Fig.~\ref{U1-configuration-2ed}(b), where circularly and linearly polarized laser beams are employed, respectively. The configuration in Fig.~\ref{U1-configuration}(a) is useful only for $U_1$, while that in Fig.~\ref{U1-configuration-2ed}(b) can also be used for $U_2$. Below, we study these two configurations separately since they involve different laser excitations.

\subsection{Qubit configuration for circularly polarized lasers}\label{sec04A}
First, we take the qubit states $|0(1)\rangle= |5s_{1/2},F=1(2),m_F=0\rangle $~\cite{Isenhower2010} to study the laser and atomic configuration for realizing $U_1$. In this case, both the lower and upper lasers should be right-hand circularly polarized, shown in Fig.~\ref{U1-configuration}. When the traps are switched off during the gate sequence, the drift of the qubits can result in phase change of the Rabi frequency, which finally leads to population leakage and phase errors as well. The larger the change of the atomic location along the propagation of the laser fields, the more the error will be. This means that if the drift speed of the qubit is set, the worst Doppler dephasing occurs when the qubit drifts along $\pm\mathbf{z}$. There are four possible worst cases characterized by the drift speeds of the control and target qubits $(\mathbf{v}_{\text{c}},\mathbf{v}_{\text{t}})=(v_z, \pm v_z )\mathbf{z}$ and $(-v_z, \pm v_z )\mathbf{z}$. Here $v_z = \sqrt{k_BT_a/m_a}$ is the root-mean-square speed of the atom along the quantization axis~($\overline{v_z^2}=\overline{v^2}/3$), and $k_B,~T_a$, and $m_a$ are the Boltzmann constant, atomic temperature, and the mass of a qubit, respectively. 

From Fig.~\ref{U1-configuration}(a), one knows that the Rabi frequencies for the control and target qubit are given by 
\begin{eqnarray}
  \Omega_{\text{c}}(t) &=& \Omega e^{i(k_{1}-k_2)z_{\text{c}}(t)}, \nonumber\\
  \Omega_{\text{t}}(t) &=& \Omega  e^{i(k_{1}-k_2)z_{\text{t}}(t)}, \label{omega_ct}
\end{eqnarray}
 respectively, where
\begin{eqnarray}
  z_{\text{c}}(t) &=&z_{\text{c}}(0)+ \mathbf{z}\cdot\mathbf{v}_{\text{c}}t,\nonumber\\
  z_{\text{t}}(t) &=&z_{\text{t}}(0)+ \mathbf{z}\cdot\mathbf{v}_{\text{t}}t.
\end{eqnarray}
The reason to assume an equal magnitude of the laser Rabi frequencies for the two qubits is as follows. We consider that the lower and upper lasers in Fig.~\ref{U1-configuration}(a) are both focused Gaussian beams~\cite{Allen1992}, with the foci at $(0,0,L/2)$ and a common radius of $\mathscr{R}=10\mu$m at their beam waist. In this case, the strengths of the laser fields at the control and target qubits are exactly the same if the two qubits are at $(0,0,0)$ and $(0,0,L)$. For qubit cooled around or below $100\mu$K, the deviation of the qubit location from the center of the dipole trap is small enough so that the strengths of the laser fields at the control and target qubits are still equal. To verify this argument, we take the setup analyzed in~\cite{Shi2018} as an example, where $\Omega_{\text{c}}(t)/  \Omega_{\text{t}}(t) $ is given by
\begin{eqnarray}
  \frac{| \Omega_{\text{c}}| }{|  \Omega_{\text{t}}| } &=& \prod_{k=1}^2\Bigg\{\frac{\mathscr{Z}_k^2+(z_{\text{t}}-\frac{L}{2})^2}{\mathscr{Z}_k^2+(z_{\text{c}}-\frac{L}{2})^2}\text{exp}\left[ \frac{\mathscr{Z}_k^2}{\mathscr{R}^2} \left( \frac{x_{\text{t}} ^2 + y_{\text{t}} ^2}{\mathscr{Z}_k^2+(z_{\text{t}}-\frac{L}{2})^2}  \right.\right.\nonumber\\
    && \left.\left.- \frac{x_{\text{c}} ^2 + y_{\text{c}} ^2}{\mathscr{Z}_k^2+(z_{\text{c}}-\frac{L}{2})^2}   \right) \right]\Bigg\},\label{RabiFluctu}
\end{eqnarray}
where the subscript $k=1$ and 2 refer to the lower and upper lasers, respectively, and $\mathscr{Z}_k=\pi \mathscr{R}^2/\lambda_k^2$. The wavelengths for the lower and upper lasers are $(\lambda_1,\lambda_2)=(795,474)$~nm. Before and after the gate sequence, we suppose that the control~(target) qubit is trapped by an optical tweezer created by a single laser beam propagating along a direction slightly tilted down~(up) from $\mathbf{z}$. By assuming the fluctuation of the qubit position along $\mathbf{x}, \mathbf{y}$, and $ \mathbf{z}$ to be within $0.3,0.3$, and $4\mu$m~\cite{Isenhower2010,Shi2018}, respectively, one can verify numerically that the average  value of $|1- | \Omega_{\text{c}}| / |  \Omega_{\text{t}}|  |$ is below $10^{-6}$. Similar calculation shows that both $|1- | \Omega_{\text{c}}| /  \Omega  |$ and $|1- | \Omega_{\text{t}}| /  \Omega  |$ are below $10^{-7}$. In the presence of qubit drift and with the trap analyzed later, $v_z$ is about $40$~nm$/\mu$s when $T_a=10\mu$K, thus the changes in qubit coordinates are around $0.1~\mu$m for a gate cycle. Therefore, the fluctuation of the qubit position along $\mathbf{x}, \mathbf{y}$, and $ \mathbf{z}$ can still be assumed to be $0.3,0.3$, and $4\mu$m, and the analysis above is still applicable. So, the deviation of the Rabi frequency from the desired one only leads to an error that is much smaller than the decay error, and can be ignored. In this case, the magnitudes of Rabi frequencies in the two qubits are approximately equal, as in Eq.~(\ref{omega_ct}).

\subsection{Qubit configuration for linearly polarized lasers}\label{sec04B}
The qubit states can also be $|0(1)\rangle= |5s_{1/2},F=1(2),m_F=1(2)\rangle $~\cite{Wilk2010}. In this case, both the lower and upper lasers should be polarized along $\mathbf{z}$, shown in Fig.~\ref{U1-configuration-2ed}. This configuration is suitable for realizing $U_2$. To use it for $U_1$, one can use a beam splitter to split one laser beam to two and use a pair of mirrors to direct them to the two qubits, shown in Fig.~\ref{U1-configuration-2ed}(a). In this configuration, we assume that before and after the gate sequence, each qubit is trapped by an optical tweezer created by a single laser beam propagating along $\mathbf{x}$. As analyzed in the text below Eq.~(\ref{RabiFluctu}), the Rabi frequencies for both the control and target qubits can be assumed equal. In the presence of atomic drift, the Rabi frequencies for the control and target qubit are given by 
\begin{eqnarray}
  \Omega_{\text{c}}(t) &=& \Omega_{\text{c0}} e^{i(k_{1}-k_2)x_{\text{c}}(t)}, \nonumber\\
  \Omega_{\text{t}}(t) &=& \Omega_{\text{t0}}  e^{i(k_{1}-k_2)x_{\text{t}}(t)}, \label{omega_ct_U2}
\end{eqnarray}
 respectively, where $\Omega_{\text{c0}}$ and $\Omega_{\text{t0}}$ are the Rabi frequencies in the absence of Doppler dephasing, and
\begin{eqnarray}
  x_{\text{c}}(t) &=& x_{\text{c}}(0) + {v}_{\text{c}}t,\nonumber\\
  x_{\text{t}}(t) &=&x_{\text{t}}(0) +{v}_{\text{t}}t,\label{motionLinerLaser}
\end{eqnarray}
where ${v}_{\text{c}},{v}_{\text{t}}=\pm\sqrt{k_BT_a/m_a}$ when the Doppler dephasing is maximal, as analyzed in Sec.~\ref{sec04A}. 

Because the configuration in Fig.~\ref{U1-configuration-2ed}(b) is useful for both $U_1$ and $U_2$, we use it for the analysis of the gate fidelity error. The protocols for $U_1$ and $U_2$ share a similarity in that either qubit is pumped only by one laser pulse. Then, the motion-induced dephasing should be similar in these two protocols. Below, we take the protocol of the $U_1$ gate as an example and analyze its fidelity. In particular, we take the parameter set 1 in Table~\ref{table1} but with reduced magnitude of $\Omega$.


\begin{table}
  \begin{tabular}{c|ccccc}
    \hline
    \hline  \text{Gate} &     $ \frac{\Omega}{2\pi}$~(MHz)&  $\frac{ \Delta}{2\pi}$ (MHz)  &$\frac{  V}{2\pi} $ (MHz) &  $L$~($\mu$m)&$t_{\text{g}}~(\mu$s) \\ \hline
$U_1$&	 0.8  & -1.54016&  2.814544 & 16.5 	 &2.30476
\\ \hline \hline
  \end{tabular}
  \caption{  \label{table3} Parameters for realizing $U_1$ according to case 1 of Table~\ref{table1} .}
  \end{table}


\subsection{ Numerical study of the gate fidelity error}\label{sec04D}
In this section, we use numerical simulation to study the fidelity of a $U_1$ gate. We consider the gate parameters in case 1 of Table~\ref{table1}, where the interaction $V$ is negative, and use the Rydberg state $|r\rangle=|100s_{1/2},m_J=1/2,m_I=3/2\rangle$ that has a positive interaction coefficient $C_6/2\pi=56.2$~THz$\mu m^6$~\cite{Saffmanr2005,Shi2014}. So, we change the signs of $\Delta$ and $V$ simultaneously for case 1 of Table~\ref{table1}, which corresponds to a similar gate with $(\beta-2\alpha)/\pi=-0.32457$. Moreover, we use a Rabi frequency $\Omega/2\pi=0.8$~MHz. This requires to shrink both $\Delta$ and $V$ by a factor of $10/0.8$. Then, a two-qubit spacing of $L= 16.5\mu$m leads to the required $V/2\pi=2.81$~MHz for this gate, see Table~\ref{table3}. 

Below, we analyze several detrimental effects that arise in the experimental realization of the gate. Among these effects, only the finite rising and falling edge of the pulse in Sec.~\ref{sec04D03} can be avoided by adjusting the pulse duration, while others should be considered in the numerical study.

\subsubsection{Degeneracy of Rydberg states}\label{sec04D01}
Population can go to states around the Rydberg state $|r\rangle$ as a leakage that cause gate errors. There is mainly one nearby Rydberg state, $|d\rangle= \zeta_1|98d_{3/2},m_J=1/2,m_I=3/2\rangle+\zeta_2 |98d_{3/2},m_J=3/2,m_I=1/2\rangle+ \zeta_3|98d_{5/2},m_J=1/2,m_I=3/2\rangle+\zeta_4 |98d_{5/2},m_J=3/2,m_I=1/2\rangle+\zeta_5 |98d_{5/2},m_J=5/2,m_I=-1/2\rangle$, that is off-resonantly coupled to the qubit state $|1\rangle$ with a detuning of about $\delta_d=1.6\times2\pi$~GHz. Here $\zeta_j$, $j=1-5$, are determined by the selection rules and satisfy the condition $\sum_j|\zeta_j|^2=1$. The next-nearest nearby states that can be coupled are $|99d_{3(5)/2},m_J,m_I\rangle$, where $m_J+m_I=2$, but with a much larger detuning of about $-5.6\times2\pi$~GHz, thus can be ignored.

The laser used for the excitation $|1\rangle\leftrightarrow|r\rangle$ can also couple the qubit state $|0\rangle$ with the Rydberg state $|s\rangle=\zeta_6|99s_{1/2},m_J=-1/2,m_I=3/2\rangle + \zeta_7|99s_{1/2},m_J=1/2,m_I=1/2\rangle$ with a detuning of $\delta_s=520\times2\pi$~MHz, where $|\zeta_6|^2+|\zeta_7|^2=1$. The next nearest Rydberg states that can be coupled with $|0\rangle$ are $|97d_{3(5)/2},m_J,m_I\rangle$, where $m_J+m_I=2$, but with a much larger detuning of about $2.1\times2\pi$~GHz. So, the main population leakage for $|0\rangle$ goes to $|s\rangle$. Details for deriving the Rabi frequencies $\Omega_d$ and $\Omega_s$ for the transitions $|1\rangle\leftrightarrow|d\rangle$ and $|0\rangle\leftrightarrow|s\rangle$ are given in Appendix~\ref{appB}. Because the leakage error is on the order of $\Omega_{d(s)}^2/\delta_{d(s)}^2$, the nearby Rydberg states will induce an error smaller than $10^{-5}$ for $\Omega/2\pi=0.8$~MHz. But if Rabi frequencies on the order of $10\times2\pi$~MHz are used, one can use lower Rydberg states with larger $\delta_{d(s)}$ so as to suppress the leakage error.

\subsubsection{Decay of Rydberg states}\label{sec04D02}
The Rydberg states $|r\rangle,~|d\rangle$, and $|s\rangle$ can decay to nearby Rydberg states and ground states. For a rubidium atom that has eight sublevels in its ground state, its Rydberg state decays into $|0(1)\rangle$ with a probability of $1/8$, and into other states with a probability of $3/4$~\cite{Zhang2012}. The former process is much less detrimental as it functions as a recycling for the qubit population. We introduce a virtual auxiliary state $|a\rangle$ that $|r\rangle$, $|d\rangle$, and $|s\rangle$ decay into with a full probability. Any population loss in $|a\rangle$ is permanent since it is not optically pumped. This model ignores that $|r\rangle$, $|d\rangle$, and $|s\rangle$ can also decay to $|0(1)\rangle$, which will slightly overestimate the gate error. The dissipative dynamics is described with the optical Bloch equation in the Lindblad form~\cite{Fleischhauer2005},
\begin{eqnarray}
  \frac{d\hat{\rho}}{dt} &=&  -i [\hat{H}',~\hat{\rho}] + \frac{1}{\tau}\sum_{k=1}^3 \left [ \hat{G}_k\hat{\rho}  \hat{G}_k^\dag - \frac{1}{2}\{\hat{G}_k^\dag \hat{G}_k,~\hat{\rho} \} \right],\nonumber\\
  \label{OpticalBloch}
\end{eqnarray}
where $\hat{\rho}$ is the density matrix of the system, $\hat{H}'$ is given by
\begin{eqnarray}
  \hat{H}'&=&\text{Eq}.~(\ref{Hamil0110})+\hat{\mathscr{H}}_c\otimes\hat{1}_t+\hat{1}_c\otimes\hat{\mathscr{H}}_t, \nonumber\\
\hat{\mathscr{H}}_{c|t}&=&  (\Omega_d|d\rangle\langle1| +\Omega_s|s\rangle\langle0| +\text{H.c.})/2,
\end{eqnarray}
where the subscript c~(t) denotes control~(target), and
\begin{eqnarray}
\hat{G}_1 &=& |a\rangle    \langle r| ,~ \hat{G}_2= |a\rangle  \langle d| ,~ \hat{G}_3= |a\rangle \langle s|  .
\end{eqnarray}
The density matrix in our model has 36 matrix elements, which is relatively large, and a solution to the optical Bloch equation can be conveniently achieved by the Monte Carlo wave-function approach~\cite{PhysRevLett.68.580}. An easy-to-use toolbox for this purpose is QuTiP~\cite{Johansson2012,Johansson2013}. Each simulation step in this method is executed by choosing a random number to see if a spontaneous decay happens. If the decay happens, the wavefunction is projected to the corresponding state determined by the random number; if it does not, the wavefunction evolves with a non-Hermitian Hamiltonian $\hat{H}'-i\sum_k\hat{G}_k^\dag \hat{G}_k/2\tau$ followed by a normalization. For small decay rates of high Rydberg states, the former process in the Monte Carlo wave-function rarely happens during the sampling. For this reason, one can also use the non-Hermitian Hamiltonian to evolve the system wavefunction, but should omit the normalization step to account for the population loss in the whole system~\cite{Petrosyan2017}.

\subsubsection{Finite rising and falling edge of laser pulse}\label{sec04D03}
The amplitude of the laser field usually has a finite rising and falling edge of width $T_{\text{edge}}$. In Ref.~\cite{Levine2018}, a two-photon Rydberg $\pi$ pulse of duration of $250$~ns was used with $T_{\text{edge}}=20$~ns. We can also assume the laser pulse has a finite rising and falling edge in our gate, each with a duration of $20$~ns. During the rising and falling, the amplitude of the Rabi frequency changes linearly between 0 and the desired magnitude. With this alteration, the pulse duration determined by Eqs.~(\ref{gatetime01}) and~(\ref{gatetime02}) no longer gives rise to the high fidelity predicted in Table~\ref{table1}. With the parameters given at the beginning of Sec.~\ref{sec04D}~(see Table~\ref{table3}), the the population loss averaged over the four input states increases to $2.5\times10^{-3}$ if we still use a pulse duration of $t_{\text{g}}=2304.76$~ns as given by Eq.~(\ref{gatetime01}). However, we can optimize the pulse duration to compensate the influence of the rising and falling of the laser pulse. Then, we find that with a gate time $t_{\text{op}}=2324.76$~ns, the population loss for the gate is only $5.14\times10^{-10}$, which is negligible. With the optimized gate duration, the angles $(\alpha,\beta,\beta-2\alpha)$ become $( -0.4502974 ,0.7748354,-0.3245698 )\pi$, which are almost identical to their values $( -0.45030 ,0.7748303,-0.3245697 )\pi$ when the pulse is square and has a duration of $t_{\text{g}}$. So, an almost perfect gate fidelity is recovered with the optimal pulse duration. For this reason, we will still assume an ideal rectangular pulse of duration $t_{\text{g}}$ in the numerical study below.


The trapezoidal pulse shape studied above is an example to demonstrate that the finite rising and falling can be compensated by adjustment of the pulse duration. In real experiments, the pulse shape depends on the parameters of the apparatus and can be different from the trapezoidal shape. Then, the value of gate time $t_{\text{op}}$ can be different from the optimal value studied above.

\subsubsection{Phase fluctuation of the laser field}\label{sec04D04}
The gate fidelity can suffer from phase noise of the laser fields~\cite{DeLeseleuc2018,Levine2018}. The natural linewidth of the laser field can be suppressed below $1$~kHz by a Pound-Drever-MogLabs lock, but the noise above the lock bandwidth is difficult to be suppressed. This can result in broad peaks in phase noise around $1$~MHz. But suppose the noise is significant, we model the power spectral density of the noise by using several discrete values of frequency $f$ according to the enhanced noise profile, i.e., the dashed curve of Fig.~7(a) in Ref.~\cite{DeLeseleuc2018}: $S_\nu(f)\approx\{3,15,25,45,70,70,10,4\}\times100~$Hz$^2/$Hz when $f = \{0.1,0.3,0.5,0.7,0.9,1,1.1,1.2\}$~MHz. Then, the random phase fluctuation of the Rabi frequency is~\cite{DeLeseleuc2018,CladePhD2004}
\begin{eqnarray}
\phi(t) &=& 2\sum_f\sqrt{S_\nu(f)/t_{\text{g}}}\cos(2\pi ft+\phi_f)/f,\label{IVC03eq01}
\end{eqnarray}
where $\phi_f$ is a random initial phase. In principle, the noise of the lower and upper lasers should be considered independently, as shown by the blue and red curves of Fig.~7(a) in Ref.~\cite{DeLeseleuc2018}. For this reason, we use the enhanced noise in Eq.~(\ref{IVC03eq01}) to represent the phase noise of the Rabi frequency.

\subsubsection{Doppler dephasing and variation of vdWI in a thermal optical trap }\label{sec04D05}

To proceed, Rabi frequencies given in Eq.~(\ref{omega_ct_U2}) should be used in the Hamiltonian $\hat{H}'$ of Eq.~(\ref{OpticalBloch}), where $\hat{H}_{v1}$ in $\hat{H}'$ should be replaced by $\hat{H}_{v2}$ defined in Eq.~(\ref{U2Hv}) for the sake of Doppler dephasing. With the motion of the qubit included as in Eq.~(\ref{motionLinerLaser}), the actual fidelity error of the entangling gate $U_1$ can be numerically studied. Because of the Doppler dephasing, the realized angles $\alpha'$ and $\beta'$ can deviate from the desired angles $\alpha$ and $\beta$ in Eq.~(\ref{gateU1}), and there can be population loss for the three input states $|01\rangle, |01\rangle $, and $|11\rangle$ as well. Then, we can use Eq.~(\ref{fidelityErrorPederson2007}) to evaluate the gate fidelity by considering the position uncertainty of the qubits and the drift of the qubits; while both of these two are related with variation of vdWI, only the latter is related with the Doppler dephasing.

In order to characterize the thermal fluctuation of the qubit positions at the beginning of each gate cycle, we give a detailed account of the optical tweezers that trap qubits before and after the gate sequence. Each optical tweezer is created by a single laser beam with wavelength $\lambda$ and waist $w$ that propagates along $x$. The parameters characterizing a trap include the trap depth $U$, the oscillation frequencies $\{\omega_x,\omega_y,\omega_z\}$, and the averaged variances $\{\sigma_x^2,\sigma_y^2,\sigma_z^2\}$ of the qubit position. The position distributions of the two qubits are modeled by Gaussian functions with variance  $\{\sigma_x^2,\sigma_y^2,\sigma_z^2\}$, where $\sigma_y^2= \sigma_z^2 =\frac{w^2}{4}\frac{T_{a}}{U}$, $\sigma_x^2=\xi^2 \sigma_y^2$, and $\xi = \sqrt2\pi w/\lambda$, where $U$ and $\xi$ are the potential depth and anisotropy factor of the trap, respectively~\cite{Saffman2005}. We set $\{w,\lambda \}=\{3.0,1.1\}\mu$m and assume an experimentally feasible trap depth of $U/k_B=20$~mK~\cite{Saffman2016} in the Monte Carlo sampling for the position distribution of qubits~\cite{Shi2017pra,Shi2017,Shi2018pra01,Shi2018prapp2}.

With the initial location of the qubits $\mathbf{r}_{\text{c}}$ and $\mathbf{r}_{\text{t}}$ determined by the above sampling method, the drift of the qubit is further considered according to Eq.~(\ref{omega_ct_U2}). In a temperature of $4.2$~K, the lifetime for $|r\rangle$ is 1.2~ms~\cite{Beterov2009}, which leads to a small decay error of about $5\times10^{-4}$ for the gate with parameters in Table~\ref{table3}. We consider two cases in Monte Carlo simulation: Maximal Doppler dephasing and maximal variation of vdWI. The first case corresponds to that both qubits drift along $\mathbf{x}$ so that the change of the phase of the Rabi frequency is maximal. The second case corresponds to that the qubits drift along $\mathbf{z}$ so that the change of vdWI is maximal.

\paragraph{Results by laser fields without phase noise.}\label{sec04D05_a}
First of all, we assume the laser fields are free from any phase noise. This is done by omiting the phase noise of Eq.~(\ref{IVC03eq01}) in the simulation. 

\begin{figure}
\includegraphics[width=3.2in]
{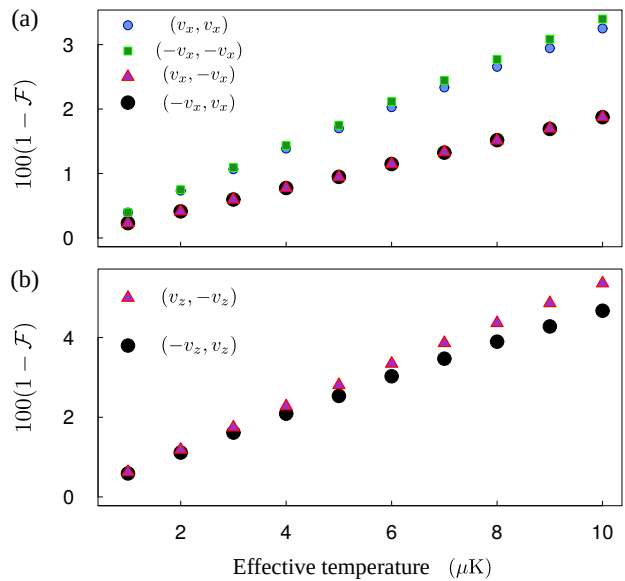}
 \caption{Error~(scale up by $100$) of the gate fidelity as a function of atomic temperature for the $U_1$ gate with $(\Omega,~V)/2\pi=(0.8,~2.81)$~MHz and $t_{\text{g}}\approx2.3~\mu$s~(see Table~\ref{table3}). In (a), four combinations of the qubit drift speeds are considered: $(\mathbf{v}_{\text{c}},\mathbf{v}_{\text{t}})=(\pm v_x, \pm v_x )\mathbf{x}$. These four cases result in maximal Doppler dephasing because the qubits drift along the direction of the laser fields. In (b), two cases are considered:$(\mathbf{v}_{\text{c}},\mathbf{v}_{\text{t}})=( v_z, - v_z )\mathbf{z}$ and $( -v_z,  v_z )\mathbf{z}$. These two cases correspond to that the two qubits approach and depart from each other, respectively, which should result in maximal variation of vdWI~\cite{Shi2018prapp2}. The error $1-\mathcal{F}$ includes Rydberg-state decay, population leakage to nearby Rydberg states, variation of vdWI, and Doppler dephasing. The thermal distribution of the initial qubit location is simulated by assuming optical dipole traps created by single lasers propagating along $\mathbf{x}$, see Fig.~\ref{U1-configuration-2ed}; the distribution is numerically simulated by Monte Carlo integration.   \label{gateerror-U2-De} }
\end{figure}

In the case of maximal Doppler dephasing, we further consider four different conditions of the qubit drift: $(\mathbf{v}_{\text{c}},\mathbf{v}_{\text{t}})=(v_x, \pm v_x )\mathbf{x}$ and $(-v_x, \pm v_x )\mathbf{x}$, where $v_x = \sqrt{k_BT_a/m_a}$ as introduced above Eq.~(\ref{omega_ct}). These four cases lead to slightly different errors in the angles $\alpha$ and $\beta$ because the change of the phases in the Rabi frequencies differs among them. Due to the slow convergence of the numerical simulation, we present results for a discrete set of atomic temperatures $T_a\in[1,10]~\mu$K in the simulation. Although an atomic temperature of several $\mu$K is quite low, it is not an unachievable value as it was demonstrated in recent Rydberg-gate experiments~\cite{Zeng2017,Picken2018}. As shown in Fig.~\ref{gateerror-U2-De}(a), the gate errors are below $10^{-2}$ when $T_a<2\mu$K, and the increase of the dephasing error is slow when the atomic temperature increases. 


In the case of maximal variation of vdWI, we only consider two conditions of the qubit drift: the two qubits approach or depart from each other along $\mathbf{z}$. These two cases correspond to the maximal variation of vdWI. Using a numerical method similar to that for Fig.~\ref{gateerror-U2-De}(a), the fidelity error was calculated and shown in Fig.~\ref{gateerror-U2-De}(b). Each gate error in Fig.~\ref{gateerror-U2-De}(b) is slightly bigger than its counterpart in Fig.~\ref{gateerror-U2-De}(a). When the six cases in Fig.~\ref{gateerror-U2-De}(a) and~\ref{gateerror-U2-De}(b) are averaged, the gate error at $T_a=1~\mu$K is about $5\times10^{-3}$. For any drift configuration different from those studied in Fig.~\ref{gateerror-U2-De}, the gate error should be bounded below the bigger value shown there. The result in Fig.~\ref{gateerror-U2-De} means that a gate fidelity larger than $0.99$ is achievable only by cooling atoms to $2\mu$K or below.

\begin{figure}
\includegraphics[width=3.2in]
{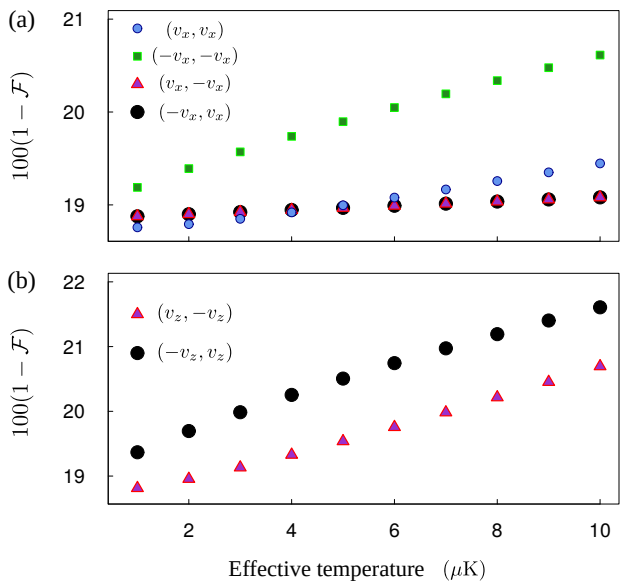}
\caption{Fidelity error~(scale up by $100$) of the $U_1$ gate with the phase noise of Eq.~(\ref{IVC03eq01}) included. Other error sources considered in Fig.~\ref{gateerror-U2-De} are incorporated with a similar numerical method, except that the random initial qubit locations are treated by averaging over several possible locations. The same parameters as in Fig.~\ref{gateerror-U2-De} are used here. Similar to Fig.~\ref{gateerror-U2-De}(a) and~\ref{gateerror-U2-De}(b), (a) and (b) correspond to the maximal Doppler dephasing and maximal variation of vdWI, respectively. Each data point is averaged over 1000 sets of $\{\phi_f\}$ in Eq.~(\ref{IVC03eq01}).  \label{gateerror-U1-noise} }
\end{figure}

\paragraph{Results by laser fields with phase noise.}\label{sec04D05_b}
Below, we show results when the Rabi frequencies have phase noise as described in Eq.~(\ref{IVC03eq01}). The time-dependent phase $\phi(t)$ in Eq.~(\ref{IVC03eq01}) will result in a much worse damping to the Rabi oscillations between the ground and Rydberg states. Because the initial term $\phi_f$ for each $f$ in $\phi(t)$ is random, it is necessary to choose enough different $\phi_f$ for the numerical simulation to converge. We found that about 1000 sets of $\phi_f$ are necessary to reach a stable simulation result when we do not consider the random distribution of the qubit locations. However, if we still use the Monte Carlo sampling for the initial locations of the two qubits, the simulation tends to run forever. Then, the random locations of the qubits at the beginning of the gate sequence are approximated by the following method: the coordinates, $(x,y,z)$, of each qubit take three possible set of values, $(\pm\sigma_x,0,0)$ and $(0,0,0)$, and the simulation result is averaged with a weighting function defined by the Gaussian trap. We use this approximation because the fluctuation along $\mathbf{y}$ and $\mathbf{z}$ is much smaller than that along $\mathbf{x}$ for the configuration of Fig.~\ref{U1-configuration-2ed}(b).

With the same set of parameters used in Fig.~\ref{gateerror-U2-De}, Fig.~\ref{gateerror-U1-noise} shows the gate errors with phase noise incorporated. In particular, Figs.~\ref{gateerror-U1-noise}(a) and~\ref{gateerror-U1-noise}(b) show the results with maximal Doppler dephasing and maximal variation of vdWI, respectively. One finds that the gate fidelity has an error of about $20\%$, an order of magnitude larger than those in Fig.~\ref{gateerror-U2-De}. This large damping is in consistent with the result shown in Fig.~7(c) of Ref.~\cite{DeLeseleuc2018} with a similar laser excitation time of about 2~$\mu$s.

\subsubsection{Prospects for high-fidelity gates}\label{sec04D06}
A comparison between Figs.~\ref{gateerror-U2-De} and~\ref{gateerror-U1-noise} shows that unless the laser noise is sufficiently suppressed, the best gate fidelity will be only $0.8$, in consistent with a recent experiment~\cite{Picken2018}. In the simulation for Fig.~\ref{gateerror-U1-noise}, we have used the enhanced spectral density in Eq.~(\ref{IVC03eq01}). But if it is reduced by several orders, for example, according to the dark gray curve of Fig.~7(a) in Ref.~\cite{DeLeseleuc2018}, the damping effect will be substantially reduced. Furthermore, The authors of Ref.~\cite{Levine2018} have recently demonstrated a suppression of such noise by more than 16 folds, thus it should be possible to achieve the gate performance shown in Fig.~\ref{gateerror-U2-De}.

Figure~\ref{gateerror-U2-De} shows that with $\Omega/2\pi=0.8$~MHz, the gate fidelity is only about $0.995$ even if $T_a=1~\mu$K, which is limited mainly by the qubit motion. The error can be suppressed by improved trapping and cooling of atoms. For example, if atoms can be cooled to their vibrational ground states~\cite{Kaufman2012,Thompson2013,Reiserer2013}, the qubits will stay very near to the trap centers, which can lead to smaller gate error caused by variation of vdWI. Furthermore, the colder the qubits are, the less severe the Doppler dephasing will be due to slow drift of qubits. However, it is nontrivial to cool atoms to very low temperatures. The main advantage of our interference method is that when the qubits are not sufficiently cooled, the gate can still attain a large fidelity. The issue with Fig.~\ref{gateerror-U2-De} is that its gate time is more than $2~\mu$s, which inevitably leads to large Doppler dephasing~\cite{DeLeseleuc2018}. Supposing a Rabi frequency of about $10\times2\pi$ is available, we found that a gate with $\mathcal{F}>0.999$ is attainable even for $T_a\approx25\mu$K~\cite{Shi2018Accuv1} if laser phase noise is suppressed, which is sufficient for fault-tolerant quantum computing based on certain strict assumptions~\cite{Knill2005}.

\section{Discussion}\label{sec05}
In this work, we have restricted our attention to two-qubit quantum gates and have shown that it is possible to construct a two-qubit entangling gate with a high fidelity in ultracold neutral atoms. Although the universal set formed by any two-qubit entangling gate and a small number of single-qubit gates is most popular~\cite{Williams2011,Nielsen2000}, it is not easy to use it for all tasks in a quantum circuit. For example, sometimes a large number of two-qubit gates are needed to construct even one three-qubit gate. For this reason, it is useful to find simple ways to realize multi-qubit gate~\cite{Isenhower2011,Shi2018prapp,Li2018,Beterov2018arX}. In principle, our method can be extended to multi-qubit gate; it is computationally more demanding to find a set of practical parameters for a high-fidelity multi-qubit gate realized with a single pulse. We leave this open for future study.

The gate examples $U_1$ and $U_2$ in this work are based on a pure excitation blockade mechanism. For neutral Rydberg atoms, however, there are a rich variety of interactions, such as the first-order dipole-dipole interaction~\cite{Ryabtsev2010,Barredo2015,Lee2017,DeLeseleuc2017} and the vdWI-based noncollinear interaction~\cite{Shi2018pra01}.  For the former type, usually multiple pairs of dipole-dipole processes exist, and it is necessary to isolate limited ones to implement a quantum gate~\cite{Ryabtsev2010,Beterov2016,Petrosyan2017,Beterov2018}. In this case, it becomes possible to use the isolated dipole-dipole flip to construct high-fidelity entangling gates, provided ground-state cooling of qubits is available~\cite{Shi2017}. Moreover, the recently found noncollinear interaction based on vdWI exhibits extra flexibility in designing Rydberg quantum gates~\cite{Shi2018pra01}. When the interference method in this work is used with the direct dipole-dipole and vdWI-based noncollinear interactions, it should be possible to find entangling gates that are more resilient to atomic motion-induced Doppler dephasing.

\section{Conclusions}\label{sec06}
We show that quantum interference in detuned Rabi oscillations of Rydberg atoms can lead to entangling gates of high intrinsic fidelity. Such gates are realized by sending to each of the qubits a single pulse of laser excitation, thus is subjected to a minimal Doppler dephasing of the transition between ground and Rydberg states. In this off-resonance interference method, the population in Rydberg states is small, so that the Rydberg-state decay error is tiny. Furthermore, the off resonance of the quantum oscillations arises not only because of vdWI, but also due to the detuning of the laser, thus the gate error due to the variation of the latter is small. Most importantly, the interference method does not require vdWI to be much smaller than the Rabi frequency, thus warrants a fast gate speed. These several advantages lead to a high fidelity in our method. Our numerical study shows that for a laser Rabi frequency below $1\times2\pi$~MHz, an entangling gate with fidelity larger than $0.99$ is possible only by cooling qubits below $2~\mu$K. Larger gate fidelity is achievable either by stronger laser powers for achieving faster gates, or by more adequate cooling. The prospect of realizing an accurate entangling Rydberg gate with minimal technical difficulty puts arrays of ultracold neutral atoms in the forefront of quantum information science.

\section*{ACKNOWLEDGMENTS}
The author thanks Yan Lu for fruitful discussions and Zhi-Wei Zhou for comments. This work was supported by the National Natural Science Foundation of China under Grant No. 11805146.

\appendix{}
  \section{Derivation of Eq.~(\ref{U1toCZ})}\label{appA}
  In this appendix, we give the details to derive Eq.~(\ref{U1toCZ}). With basis $\{|0\rangle,|1\rangle\}$ for the control or target qubit, we define the phase shift gate
\begin{eqnarray}
  \hat{\mathscr{P}}_\phi  &=& \left( \begin{array}{cc}
    1& 0 \\
    0 & e^{i\phi}
    \end{array} \right).
\end{eqnarray}
 Define the following phase change on the qubits:
\begin{eqnarray}
\hat{O}_1&=&[  \hat{\mathscr{P}}_{-\beta/2}]_{\text{c}}\otimes[ \hat{\mathscr{P}} _{-\alpha} ]_{\text{t}},\nonumber\\
\hat{O}_2&=&[ \hat{\mathscr{P}}_{-\beta}]_{\text{c}}\otimes[  \hat{\mathscr{P}}_{-2\alpha} ]_{\text{t}},\label{O1definition}
\end{eqnarray}
where the operator left~(right) of `$\otimes$' operates on the control~(target) qubit. The above gates change Eq.~(\ref{gateU1}) to
\begin{eqnarray}
\hat{O}_1 U_1&=&|0\rangle\langle0|\otimes \hat{I} + |1\rangle\langle1|\otimes e^{i(\alpha-\beta/2)\sigma_3}.
\end{eqnarray}
and $U_1^2$ to
\begin{eqnarray}
\hat{O}_2 U_1^2&=&|0\rangle\langle0|\otimes \hat{I} + |1\rangle\langle1|\otimes e^{i(2\alpha-\beta)\sigma_3}.\label{App_EQ01}
\end{eqnarray}

  In order to construct a CNOT from Eq.~(\ref{sec02D_EQ01}), $U_1$~(with the case 1 of Table~\ref{table1}), and single-qubit gates, we need the following gate
  \begin{eqnarray}
\hat{G}_1&=&|0\rangle\langle0|\otimes \hat{I} + |1\rangle\langle1|\otimes e^{[-i\pi/2-i(2\alpha-\beta)]\sigma_3},
\end{eqnarray}
  so that $\hat{G}_1\hat{O}_1 U_1^2$ becomes $|0\rangle\langle0|\otimes \hat{I} + |1\rangle\langle1|\otimes e^{-i\pi\sigma_3/2}$. $\hat{G}_1$ can be derived by following exercise 4.15 of Ref.~\cite{Nielsen2000}. Because $|\pi/2+(2\alpha-\beta)|<|2\alpha-\beta|$, $\hat{G}_1$ is formed by using two $U_1$ when assisted by appropriate single-qubit gates $\hat{A}$ and $\hat{B}$:
   \begin{eqnarray}
\hat{G}_1&=&[\hat{I} \otimes \hat{B}][\hat{O}_1U_1 (\hat{I} \otimes \hat{A})\hat{O}_1 U_1  (\hat{I} \otimes \hat{A}^\dag) ][\hat{I} \otimes \hat{B}^\dag],\nonumber
\end{eqnarray} 
where the conjugation of $\hat{A}$ in $\hat{O}_1U_1$ rotates the axis of the rotation operator $\hat{O}_1U_1$, while that of $\hat{B}$ rotates the axis back to $\mathbf{z}$. To solve $\hat{A}$ and $\hat{B}$, we start from the fact that $[\hat{O}_1U_1 (\hat{I} \otimes \hat{A})\hat{O}_1 U_1  (\hat{I} \otimes \hat{A}^\dag) ]$ must represent a controlled rotation of angle $-2[-\pi/2-(2\alpha-\beta)]$ about an axis $\mathbf{n}_{12}$, and $(\hat{I} \otimes \hat{A})\hat{O}_1 U_1  (\hat{I} \otimes \hat{A}^\dag)$ is a rotation of $\beta-2\alpha$ around $\mathbf{n}_2$, where the rotation axis $\mathbf{n}_2$ and $\mathbf{n}_{12}$ are to be solved. By using Eqs.~(4.21) and (4.22) of Ref.~\cite{Nielsen2000} we have
  \begin{eqnarray}
    \cos\left(\frac{\pi}{2}+2\alpha-\beta\right) &=& \cos^2\left(\frac{\beta}{2}-\alpha\right) \nonumber\\
    &&- \sin^2\left(\frac{\beta}{2}-\alpha\right) \mathbf{n}_2\cdot\mathbf{z},\label{App01}\\
    \sin\left(\frac{\pi}{2}+2\alpha-\beta\right) \mathbf{n}_{12}&=&\frac{1}{2} \sin\left(\beta-2\alpha\right)( \mathbf{n}_2+\mathbf{z} )\nonumber\\
    &&- \sin^2\left(\frac{\beta}{2}-\alpha\right) \mathbf{n}_2\times\mathbf{z}.\label{App02}
\end{eqnarray} 
  The solution of $\mathbf{n}_2$ to Eq.~(\ref{App01}) is not unique. If $\beta/2-\alpha\in(-\pi,\pi)/2$ and $\beta/2-\alpha\notin(-3\pi/8,\pi/8)$, we can choose $\mathbf{n}_2= a\mathbf{z} +\sqrt{1- a^2}\mathbf{x}$, where
    \begin{eqnarray}
       a&=&   \frac{ \cos^2\left(\frac{\beta}{2}-\alpha\right) + \sin\left(2\alpha-\beta\right)}{\sin^2\left(\frac{\beta}{2}-\alpha\right)}.
\end{eqnarray} 
    Substitution of the above result into Eq.~(\ref{App02}) leads to
    \begin{eqnarray}
      \mathbf{n}_{12}&=& \frac{\sin\left(\beta-2\alpha\right) \sqrt{1- a^2}  }{2\cos\left(2\alpha-\beta\right)}  \mathbf{x} \nonumber\\
      && + \frac{ \sin^2\left(\frac{\beta}{2}-\alpha\right) \sqrt{1- a^2}  }{\cos\left(2\alpha-\beta\right)}  \mathbf{y} \nonumber\\
      && + \frac{\sin\left(\beta-2\alpha\right) (1+ a)  }{2\cos\left(2\alpha-\beta\right)}  \mathbf{z}. \nonumber
\end{eqnarray} 
    
    With $\mathbf{n}_{2}$ and $\mathbf{n}_{12}$ at hand, we can solve $\hat{A}$ and $\hat{B}$. From the fact that $(\hat{I} \otimes \hat{A})\hat{O}_1 U_1  (\hat{I} \otimes \hat{A}^\dag)$ is a rotation of $\beta-2\alpha$ around $\mathbf{n}_2$ and by using Eq.~(4.8) of Ref.~\cite{Nielsen2000}, we have $\hat{A}\sigma_3 \hat{A}^\dag = a\sigma_3 +\sqrt{1- a^2}\sigma_1 $. This leads to $\hat{A}=e^{-i\sigma_2\arccos a/2 }$. Similarly, by using the equation $\hat{B}   [ \mathbf{n}_{12} \cdot(\sigma_1\mathbf{x},\sigma_2\mathbf{y},\sigma_3\mathbf{z}) ]\hat{B}^\dag = \sigma_3  $, we obtain, 
    \begin{eqnarray}
\hat{B} &=& e^{-i\theta_3 \sigma_3/2}  e^{-i\theta_2 \sigma_2/2} ,     
\end{eqnarray} 
    where $\theta_2$ and $\theta_3$ are given by 
    \begin{eqnarray}
      \theta_2 &=& \arccos \frac{\sin\left(\beta-2\alpha\right) (1+ a)  }{2\cos\left(2\alpha-\beta\right)},\nonumber\\
      \theta_3&=& \arctan\frac{2 \sin^2\left(\frac{\beta}{2}-\alpha\right)   }{\sin\left(\beta-2\alpha\right)} .
\end{eqnarray} 

In conclusion, with the $U_1$ gate of case 1 in Table~\ref{table1}, we need four $U_1$ and several single-qubit gates to construct the controlled $\pi$-rotation about the $\mathbf{z}$ axis, i.e., 
    \begin{eqnarray}
  &&    |0\rangle\langle0|\otimes \hat{I} + |1\rangle\langle1|\otimes e^{-i\pi\sigma_3/2} \nonumber\\
&&   =   [\hat{I} \otimes \hat{B}][\hat{O}_1U_1 (\hat{I} \otimes \hat{A})\hat{O}_1 U_1  (\hat{I} \otimes \hat{A}^\dag) ][\hat{I} \otimes \hat{B}^\dag]\hat{O}_2 U_1^2.\nonumber\\
      \label{Appeq10}
\end{eqnarray}   
If we further use a phase gate $\hat{\mathscr{P}}_{\pi/2}$ on the control qubit, one can construct a C$_{\text{Z}}$, i.e., $[\hat{\mathscr{P}}_{\pi/2}\otimes\hat{I}]\cdot$[Eq.~(\ref{Appeq10})]$=$C$_{\text{Z}}$. To realize a CNOT, we need two extra rotations of $\pm\pi/2$ about $\mathbf{y}$ in the target, i.e., $[\hat{\mathscr{P}}_{\pi/2}\otimes e^{-i\pi\sigma_2/4}  ]$[Eq.~(\ref{Appeq10})]$[\hat{I}\otimes e^{i\pi\sigma_2/4}  ]$=CNOT.

 \section{ Rabi frequency for the population-leaking channel   }\label{appB}
For the diagram in Fig.~\ref{U1-configuration-2ed}, the transition between $|1\rangle=5s_{1/2},F=1(2),m_F=1(2)$ and $|r\rangle$ has a Rabi frequency 
\begin{eqnarray}
  \Omega_0 \approx\left\{ \left[\frac{\Omega_{\text{low}}^{(3)}\Omega_{\text{upp}} ^{(3)}}{2\delta_{\text{2-pho}}}\right]^2+\left[\frac{\Omega_{\text{low}}^{(2)}\Omega_{\text{upp}} ^{(2)}}{2(\delta_{\text{2-pho}}+\delta_{\text{Hyp}})}\right]^2\right\}^{1/2},\nonumber\\
  \label{appenEq01}
\end{eqnarray}
where $\Omega_{\text{low}}^{(j)}$ and $\Omega_{\text{upp}}^{(j)}$ are the Rabi frequencies of the lower transition $|1\rangle\leftrightarrow|e\rangle$ and upper transition $|e\rangle\leftrightarrow|r\rangle$ via the $F=j$ hyperfine level of the $5P_{3/2}$ state, respectively. Here $\delta_{\text{2-pho}}$ is a GHz-scale detuning for the lower transition via the $|5P_{3/2},F=2\rangle$ state, and $\delta_{\text{Hyp}}$ is the energy difference between the $F=2$ and $F=3$ hyperfine states, which is only $267\times2\pi$~MHz for $^{87}$Rb~\cite{Steck2015}. According to the Wigner-Eckart theorem, 
\begin{eqnarray}
  \Omega_{\text{low}}^{(2)} &=&-e \mathscr{E}_{\text{low}}(5P_{3/2},F=2 ||r|| 5S_{1/2},F=2) C_{2,0,2}^{2,1,2} \nonumber\\ &=&e \mathscr{E}_{\text{low}}(5P_{3/2} ||r|| 5S_{1/2})  C_{2,0,2}^{2,1,2} \nonumber\\ &&\times \sqrt{10} \left\{\begin{array}{ccc}1/2& 3/2& 1\\
  2&2&3/2\end{array} \right\} , \nonumber\\
\Omega_{\text{upp}}^{(2)} &=&-e \mathscr{E}_{\text{upp}}(5P_{3/2} ||r|| nS_{1/2}) C_{1/2,0,1/2}^{1/2,1,3/2} C_{1/2,3/2,2}^{3/2,3/2,2} \nonumber\\ &=&e \mathscr{E}_{\text{upp}}(5P ||r|| nS)  C_{1/2,0,1/2}^{1/2,1,3/2} C_{1/2,3/2,2}^{3/2,3/2,2} \nonumber\\ &&\times \sqrt6 \left\{\begin{array}{ccc}1& 0& 1\\
  1/2&3/2&1/2\end{array} \right\},\nonumber\\
  \Omega_{\text{low}}^{(3)} &=&-e \mathscr{E}_{\text{low}}(5P_{3/2},F=3 ||r|| 5S_{1/2},F=2) C_{2,0,2}^{2,1,3} \nonumber\\ &=&-e \mathscr{E}_{\text{low}}(5P_{3/2} ||r|| 5S_{1/2}) C_{2,0,2}^{2,1,3} \nonumber\\ &&\times \sqrt{14} \left\{\begin{array}{ccc}1/2& 3/2& 1\\
  3&2&3/2\end{array} \right\} , \nonumber\\
\Omega_{\text{upp}}^{(3)} &=&-e \mathscr{E}_{\text{upp}}(5P_{3/2} ||r|| nS_{1/2}) C_{1/2,0,1/2}^{1/2,1,3/2} C_{1/2,3/2,2}^{3/2,3/2,3} \nonumber\\ &=&e \mathscr{E}_{\text{upp}}(5P ||r|| nS)  C_{1/2,0,1/2}^{1/2,1,3/2} C_{1/2,3/2,2}^{3/2,3/2,3} \nonumber\\ &&\times \sqrt6 \left\{\begin{array}{ccc}1& 0& 1\\
  1/2&3/2&1/2\end{array} \right\},\label{appenEq02}
\end{eqnarray}
where $n=100$, the doubled bars indicate the reduced matrix element, $e$ is the elementary charge, $C$ is a Clebsch-Gordan coefficient, $\{\cdots\}$ is a 6-j symbol, and $\mathscr{E}$ is the field strength.

To study the Rabi frequency from $|1\rangle$ to $|d\rangle= \zeta_1|98d_{3/2},m_J=1/2,m_I=3/2\rangle+\zeta_2 |98d_{3/2},m_J=3/2,m_I=1/2\rangle+ \zeta_3|98d_{5/2},m_J=1/2,m_I=3/2\rangle+\zeta_4 |98d_{5/2},m_J=3/2,m_I=1/2\rangle+\zeta_5 |98d_{5/2},m_J=5/2,m_I=-1/2\rangle$, we split $|d\rangle$ to two unnormalized parts, $|d\rangle= |d_{1}\rangle+|d_{2}\rangle$, where $|d_{1}\rangle= \zeta_1|98d_{3/2},m_J=1/2,m_I=3/2\rangle+\zeta_2 |98d_{3/2},m_J=3/2,m_I=1/2\rangle$. The Rabi frequency to $|d_{1}\rangle$ can be represented by
\begin{eqnarray}
 \Omega_{d1}= ( \zeta_1^2\Omega_{d1-1}^2+\zeta_2^2\Omega_{d1-2}^2)^{1/2},\label{appenEq03}
\end{eqnarray}
where $\zeta_1$ and $\zeta_2$ are given by the selection rules, and $\Omega_{d1-1}$ and $\Omega_{d1-2}$ denote the transition frequency to $|98d_{3/2},m_J=1/2,m_I=3/2\rangle$ and $ |98d_{3/2},m_J=3/2,m_I=1/2\rangle$, respectively, both of which can be derived in a similar way shown in Eqs.~(\ref{appenEq01}) and (\ref{appenEq02}). When $\delta_{\text{2-pho}}$ is known, this method gives us an $\Omega_0$ that is a function of $\mathscr{E}_{\text{low}}\mathscr{E}_{\text{upp}} (5P_{3/2} ||r|| 5S_{1/2}) (5P ||r|| nS)$, and $\Omega_{d1-1}$ and $\Omega_{d1-2}$ are functions of $\mathscr{E}_{\text{low}}\mathscr{E}_{\text{upp}} (5P_{3/2}||r|| 5S_{1/2}) (5P ||r|| nD)$. One can evaluate $|  (5P ||r|| 98D)/   (5P ||r|| 100S)|$ by resorting to the the ``open-source library for calculating properties of alkali Rydberg atoms''~\cite{Sibalic2017}, or directly by using the semiclassical approach in Ref.~\cite{Kaulakys1995}. The Rabi frequency, denoted by $\Omega_{d2}$, from $|1\rangle$ to $|d_{2}\rangle= \zeta_3|98d_{5/2},m_J=1/2,m_I=3/2\rangle+\zeta_4 |98d_{5/2},m_J=3/2,m_I=1/2\rangle+\zeta_5 |98d_{5/2},m_J=5/2,m_I=-1/2\rangle$ can be derived in a similar manner. Then, the Rabi frequency from $|1\rangle$ to $|d\rangle$ is given by $\Omega_d=\sqrt{\Omega_{d1}^2 + \Omega_{d2}^2}$.

To estimate the Rabi frequency, $\Omega_s$, for the leaking channel $|0\rangle\rightarrow|s\rangle=\zeta_5|99s_{1/2},m_J=-1/2,m_I=3/2\rangle + \zeta_6|99s_{1/2},m_J=1/2,m_I=1/2\rangle$, where $|\zeta_5|^2+|\zeta_6|^2=1$, we note that the energy difference between the the $F=2$ and $F=1$ hyperfine states is only $157\times2\pi$~MHz. So, all the three hyperfine levels $F=1-3$ of the $5P_{3/2}$ state contribute to population leakage. Using a similar method shown above, one can derive a Rabi frequency $\Omega_s$ that is also a function of $\mathscr{E}_{\text{low}}\mathscr{E}_{\text{upp}} (5P_{3/2} ||r|| 5S_{1/2}) (5P ||r|| 99S)$, where $| (5P ||r|| 99S)/   (5P ||r|| 100S)|$ can be calculated in a similar way for calculating $|  (5P ||r|| 98D)/   (5P ||r|| 100S)|$. For $n=100$ and $\delta_{\text{2-pho}}/2\pi=2$~GHz, we found $\Omega_d:\Omega_0:\Omega_s\approx2:1:0.84$. When this relation is used in the numerical simulation, the phases of the Rabi frequencies $\Omega_d$ and $\Omega_s$ are also subject to the Doppler dephasing as discussed in Sec.~\ref{sec04D05}.

%


\end{document}